\documentclass[10pt,letterpaper]{article}
\usepackage[top=0.85in,left=2.75in,footskip=0.75in]{geometry}

\usepackage{amsmath,amssymb}

\usepackage{changepage}

\usepackage[utf8x]{inputenc}

\usepackage{textcomp,marvosym}

\usepackage{cite}

\usepackage{nameref,hyperref}

\usepackage{microtype}
\DisableLigatures[f]{encoding = *, family = * }

\usepackage[table]{xcolor}

\usepackage{array}

\newcolumntype{+}{!{\vrule width 2pt}}

\newlength\savedwidth

\raggedright
\usepackage[aboveskip=1pt,labelfont=bf,labelsep=period,justification=raggedright,singlelinecheck=off]{caption}

\makeatletter
\renewcommand{\@biblabel}[1]{\quad#1.}
\makeatother

\usepackage{lastpage,fancyhdr,graphicx}
\usepackage{epstopdf}
\fancyheadoffset[L]{2.25in}
\fancyfootoffset[L]{2.25in}
\RequirePackage[OT1]{fontenc}
\RequirePackage{amsthm,amsmath}
\usepackage{amsfonts}
\RequirePackage[authoryear]{natbib}
\newcommand{\X}{{\mathbf{X}}}

\newcommand{\bbeta}{\boldsymbol{\beta}}

\begin{document}

\begin{center}

\phantom{yoyoyoyo}

\vspace{2cm}

{\LARGE {\bf Community vibrancy and its relationship}} 

\vspace{0.25cm}

{\LARGE {\bf with safety in Philadelphia}}

\vspace{2cm}

Wichinpong Park Sinchaisri$^{\, 1 \, }$ and Shane T. Jensen$^{\, 2 \, * \, }$

\end{center}

\vspace{2cm}

\noindent
$^{1}$ Department of Operations, Information, and Decisions, The Wharton School, University of Pennsylvania, Philadelphia, Pennsylvania, United States of America \\

\noindent
$^{2}$ Department of Statistics, The Wharton School, University of Pennsylvania, Philadelphia, Pennsylvania, United States of America \\

\vspace{1cm}

\noindent
$^*$ Corresponding author \\

\noindent
E-mail: stjensen@wharton.upenn.edu (STJ)

\vspace{2cm}

\noindent
Both authors contributed equally to this work.

\thispagestyle{empty}

\newpage 

\setcounter{page}{0}
\pagenumbering{arabic}
\setcounter{page}{1}

\begin{flushleft}

{\LARGE {\bf Community vibrancy and its relationship}} 

\vspace{0.25cm}

{\LARGE {\bf with safety in Philadelphia}}

\vspace{1cm}

{\bf Wichinpong Park Sinchaisri and Shane T. Jensen}

\end{flushleft}

\vspace{1cm}

%\bigskip

\section*{Abstract}

To what extent can the strength of a local urban community impact neighborhood safety?  We construct measures of {\it community vibrancy} based on a unique dataset of block party permit approvals from the City of Philadelphia.   Our first measure captures the overall volume of block party events in a neighborhood whereas our second measure captures differences in the type (regular versus spontaneous) of block party activities.    We use both regression modeling and propensity score matching to control for the economic, demographic and land use characteristics of the surrounding neighborhood when examining the relationship between crime and our two measures of community vibrancy.   We conduct our analysis on aggregate levels of crime and community vibrancy from 2006 to 2015 as well as the trends in community vibrancy and crime over this time period.  We find that neighborhoods with a higher number of block parties have a significantly higher crime rate, while those holding a greater proportion of spontaneous block party events have a significantly lower crime rate.   We also find that neighborhoods which have an increase in the proportion of spontaneous block parties over time are significantly more likely to have a decreasing trend in total crime incidence over that same time period.

%\linenumbers

\section{Introduction}\label{introduction}

Why does the crime rate vary so strikingly between neighborhoods in large cities?  Common factors associated with high crime rates include poverty levels, job availability, policing, and the average age of the population.  A theory proposed by \cite{ShaMcK42} first connected these community characteristics with crime rates through \emph{social disorganization}: disadvantaged neighborhoods facing poverty, cultural differences, and high residential mobility generally struggle to develop strong bonds among their members and tend to have high delinquency rates.  

Since then, this model has been tested empirically by several researchers.  
\cite{crutchfield1982crime} found that high rates of mobility negatively affected social integration, lowering the effectiveness of community informal control mechanisms. \cite{SamGro89} demonstrated that between-community variations in social disorganization significantly affected crime rates. These findings seem to suggest that the nature of social interaction within a community is correlated with local safety.   

In this paper, we investigate the association between crime incidence and measures of community social cohesion or {\bf vibrancy} based on the tradition of block parties in the city of Philadelphia.   We create quantitative measures of community vibrancy in local areas of the city of Philadelphia using a dataset of block party permit approvals.  Since 75\% of a street's residents need to agree to hold a block party, this data provides a unique perspective on the cohesion of local communities within a large and diverse urban environment.

%\begin{center}
%\textcolor{red}{START NEW FOR LIT REVIEW}
%\end{center}

\cite{DeaHilCha15} found that block parties were associated with increased bonding social capital in Black neighborhoods in Philadelphia.  They also suggested that block parties might be reflective of {\bf collective efficacy}, the willingness of residents to intervene as guardians on behalf of the community \citep{SamRauEar97}.    

Many theories in criminology suggest that collective efficacy and guardianship within local communities are important for crime prevention.  Situational crime prevention connects guardianship to the ease of different types of criminal activity \citep{Cla95,WilCul18}.   Human territorial functioning \citep{Tay88} and broken windows theory \citep{WilKel82} suggest that crime is fostered in locations that lack guardianship and public displays of community responsibility.

Empirical studies also support that collective efficacy and guardianship within a community are associated with reductions in crime.  \cite{SamGro89} found that variations in neighborhood cohesion and community participation could explain different rates of criminal victimization and conviction in British localities.
\cite{bellair1997social} explored the consequences of frequent and infrequent interaction among neighbors and finds that the type of interaction matters.  Getting together once a year or more with neighbors has the most consistent and generally strongest effect on burglary, motor vehicle theft, and robbery. 

\cite{MarBelLis01} built on the earlier social disorganization theories of \cite{ShaMcK42} by using several waves of the British Crime Survey to demonstrate that decreases in neighborhood cohesion can lead to increases in crime, disorder and fear which further decreases neighborhood cohesion.  This feedback between community cohesion and disorder was also observed in the longitudinal study of \cite{SteHip11}.   \cite{WicHipSar13} examined the contribution of social ties and perceived social cohesion for the development of collective efficacy norms in Australian communities.  However, as \cite{HipWo15} argued, constructs such as collective efficacy or community cohesion are subtle and difficult to directly observe.  

In this paper, we use the term {\bf community vibrancy} to reflect observable public displays of community cohesion and social bonding that the aforementioned studies suggest should be associated with neighborhood safety.  We create two quantitative measures of community vibrancy that are intended to capture different aspects of community cohesion and social organization: the total number of block party events and the proportion of spontaneous block party events in a neighborhood. This second measure distinguishes between two major types of block party events: regular block party events for public or religious holidays versus spontaneous block party events.    

These two types of block party events could reflect different types of community cohesion as regular block party events are more likely to build upon established institutions (e.g. churches) whereas spontaneous block party events are more likely to be organized around specific events that reflect the dynamics and cohesion among individual community members.   

These different types of events could also signal different levels (or types) of collective efficacy and guardianship, potentially explaining variation in the prevention of crime.   Regular block party events could be indicative of a central religous or neighborhood organization that facilitates strong but diffuse cohesion across a large proportion of the community.   In contrast, spontaneous block party events such as birthdays or graduations may be more indicative of the role that particular individuals or households have in organizing public events within a particular community.

We then investigate whether there is an association between these measures of community vibrancy and crime incidence at the neighborhood level in Philadelphia.  We also examine the relationship between changes in community vibrancy over time and trends in crime over time.  

However, these relationships are potentially confounded by many other neighborhood factors that are also related to either our created measures of community vibrancy or crime incidence.  For example, \cite{wu2018urban} defined neighborhood vibrancy using a GPS-based activity survey in suburban Beijing and found that high density and mixed land use were positively correlated with neighborhood vibrancy.  

To address this possibility, we incorporate data on the economic, demographic and land use characteristics of Philadelphia neighborhoods into our analyses.  We use two statistical techniques, regression modeling and propensity score matching, to estimate the association between crime and community vibrancy while controlling for these other neighborhood factors.  

%Our results demonstrate that neighborhoods with a larger number of block parties per year have a significantly higher annual crime rate. However, those holding a greater proportion of spontaneous block party events have a significantly lower crime rate. We also find that neighborhoods with an increase in the ratio of spontaneous block parties over time, potentially indicating a stronger sense of community, are significantly more likely to have a decreasing trend in total crime incidence over that same time period. 

%This paper is organized as follows. We use Philadelphia block party permit data to define measures of community vibrancy in Section~\ref{vibrancy}. We explore the relationship between our community vibrancy measures and other neighborhood characteristics (economic, demographic, and land use) in Section~\ref{othercharacteristics}. We examine crime incidence at the neighborhood level in Philadelphia in Section~\ref{crimedata} and then investigate the association between overall crime incidence and our community vibrancy measures in Section~\ref{crime-aggregate}. We also investigate the relationship between changes over time in crime incidence and community vibrancy in Section~\ref{trends}. We summarize our findings in Section~\ref{discussion}.

\section{Measuring Community Vibrancy through Block Parties} \label{vibrancy}

\subsection{Data on Block Parties in Philadelphia}\label{data}

Our dataset contains 68,553 permit approvals for a block party across 10,347 unique locations (by street address) in the city of Philadelphia from January 2006 to May 2016.  This data was made available to us by the author of \cite{geeting2016} and can be accessed at {\bf link withheld during review}.   All permits in this data are for one-day events, although we do observe that some blocks organize events on consecutive days.  Since we do not observe the full details of the event nor its planner, we consider events on consecutive days as separate events.

In this paper, we study community vibrancy at the neighborhood level of resolution.  We will define our neighborhood units as the ``block group" geographical units established by the US Census Bureau.  There are 1,336 US Census block groups in the city of Philadelphia.   These US census block groups consist of 10-20 city blocks which generally matches our concept of a ``neighborhood", and the block group level is the highest resolution at which the US Census Bureau publicly releases economic data.   We aggregate the 68,553 block party permits within these 1,336 neighborhoods in Philadelphia.  

There are 30 unique event types for these block party permits which we group into two main categories: {\it regular} events such as national or religious holidays versus {\it spontaneous} events that are not tied to a regular holiday.  The breakdown of the event types within these two categories is:

\begin{itemize}
\item {\bf Regular events} (7.5\%)
\begin{itemize}
\item \textit{Public holiday}: 4th of July, Labor Day, Memorial Day, New Year's Day, New Year's Eve, May Day, Christmas Party, Father's Day, Mother's Day, Halloween Party (7.3\%)
\item  \textit{Religious}: Church Service, Communion, Religious Event (0.2\%)
\end{itemize}
\item {\bf Spontaneous events} (92.5\%)
\begin{itemize}
\item  \textit{Community}: Community Fun Day, Easter Egg Hunt, National Night Out, Prom, Spring Festival, Arts \& Crafts Show, Health Fair, Stop The Violence Crusade, Dedication, Serenade (92.1\%)
\item  \textit{Personal}: Baby Shower, Birthday Party, Graduation Party, Repass, Wedding Reception, Wedding (0.4\%)
\end{itemize}
\end{itemize}

As we see above, regular block party events are associated with public or religious holidays and are more likely to build upon established institutions within the community such as churches or neighborhood organizations.   In contrast, spontaneous block party events are more likely to be organized by specific persons or households and could be more reflective of the dynamics among individual community members.  In designing our two measures of community vibrancy, we attempt to capture both the overall volume of community activity in a neighborhood with our first measure as well as distinguish between regular versus spontaneous types of community activities with our second measure.   

As we discuss in our introduction, these two different types of block party activities could relate to different types of community cohesion and hence have different relationships with crime prevention.   The total number of block party events (both regular and spontaneous) is a measure of the overall strength of community cohesion and potentially general guardianship against crime.   Within this total amount of block party activity, higher versus lower spontaneous proportion may provide additional information on the scale of community involvement in these events, with spontaneous events (like birthdays and graduations) likely to be more focused around particular households.  The more concentrated nature of these spontaneous events could result in more localized contributions to community cohesion and guardianship against crime.

\subsection{Community Measure 1: Total Number of Block Party Events}\label{totalnumber}

We first consider the total number of regular or spontaneous block party events held within each neighborhood.  The total number of block party events held in a particular neighborhood is a simple and intuitive measure of the community vibrancy of that neighborhood.  In Fig S1 in our supplementary materials, we show the total number of block party events within each neighborhood of Philadelphia, aggregated across the entire time span of our data (2006-2016).  

We find that neighborhoods that have the largest total number of block party events are in the North Philadelphia area.  West Philadelphia and South Philadelphia also have several neighborhoods with a large total number of block party events, whereas the outlying suburban communities in the Northwest and Northeast parts of the city have relatively few total number of block party events.   We will examine the trend over time in the total number of block party events aggregated by year in Section~\ref{vibrancytimetrends} below.  

\subsection{Community Measure 2: Spontaneous Proportion}\label{spontaneity}

In addition to the total number of block party events held in each neighborhood, we are also interested in the distinction between spontaneous versus regular block party events, as outlined in Section~\ref{data} above.  For each neighborhood in Philadelphia, we compute the proportion of the number of spontaneous events to the total number of events (spontaneous or regular).  

This {\it spontaneous proportion} is generally quite high since over 90\% of block party events are categorized as spontaneous.   Almost all neighborhoods (97.5\%) have a spontaneous proportion above 0.8, but there is still considerable variation in spontaneous proportion between neighborhoods.  In Fig S1 in our supplementary materials, we show the spontaneous within each neighborhood of Philadelphia, aggregated across the entire time span of our data (2006-2016).

While North Philadelphia contains the neighborhoods with the largest total number of block party events, we see that these North Philadelphia neighborhoods also have a lower spontaneous proportion than other areas of the city. Center city and the Northwest and Northeast suburban communities contain the neighborhoods with the highest spontaneous proportions in Philadelphia.   We will also examine the trend over time in the spontaneous proportion in Section~\ref{vibrancytimetrends} below. 

\subsection{Trends over Time in Community Measures} \label{vibrancytimetrends}

%Figure~\ref{vibrancymeasures} shows the spatial distribution of our two community measures that have been aggregated over the entire time span of our data (2006-2016).   However, we can also examine how each of these community measures varies over that same period, both in terms of monthly and yearly trends.  

\begin{figure}[!h]
\centering
\includegraphics[width=0.49\textwidth]{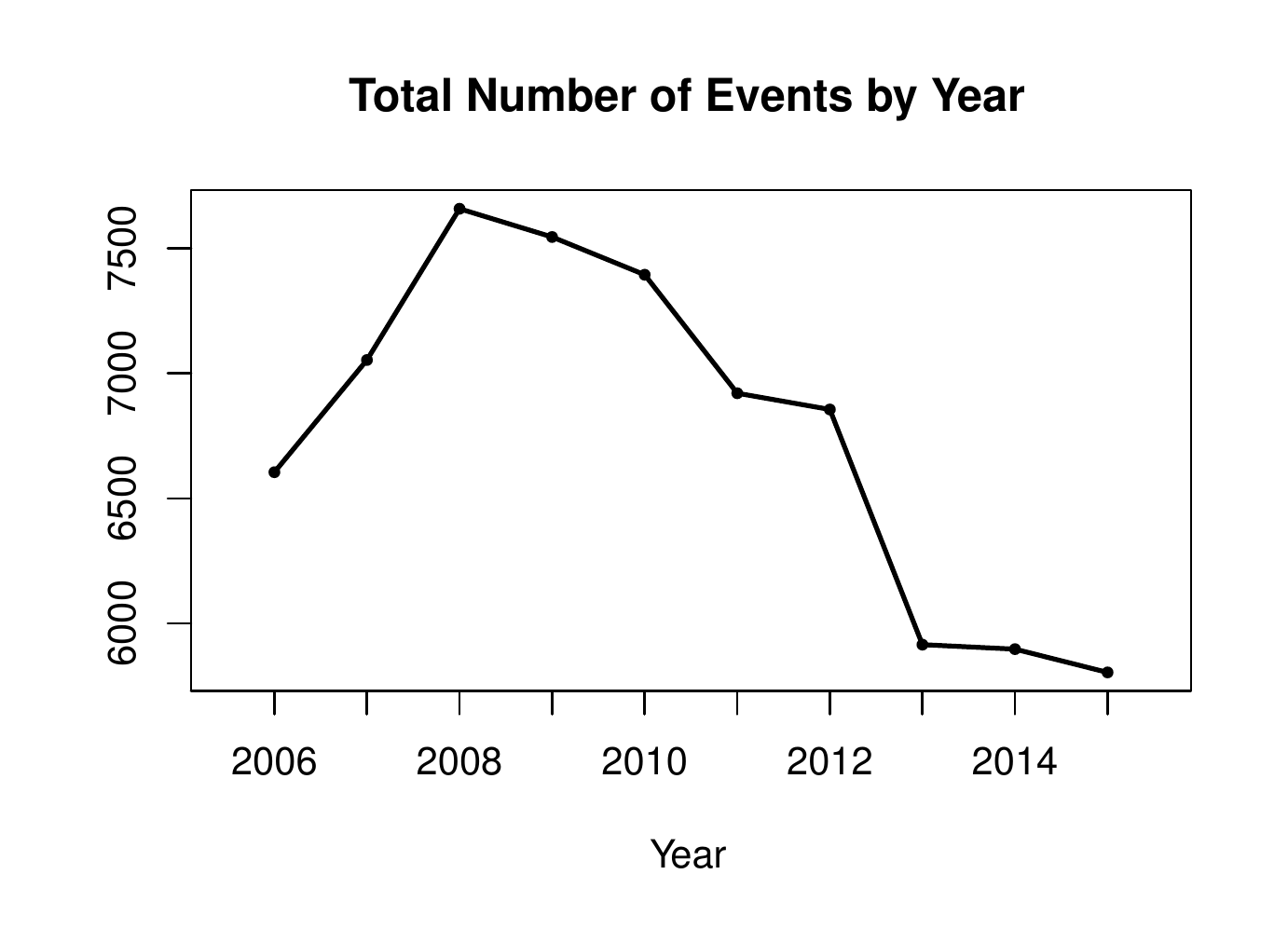}
\includegraphics[width=0.49\textwidth]{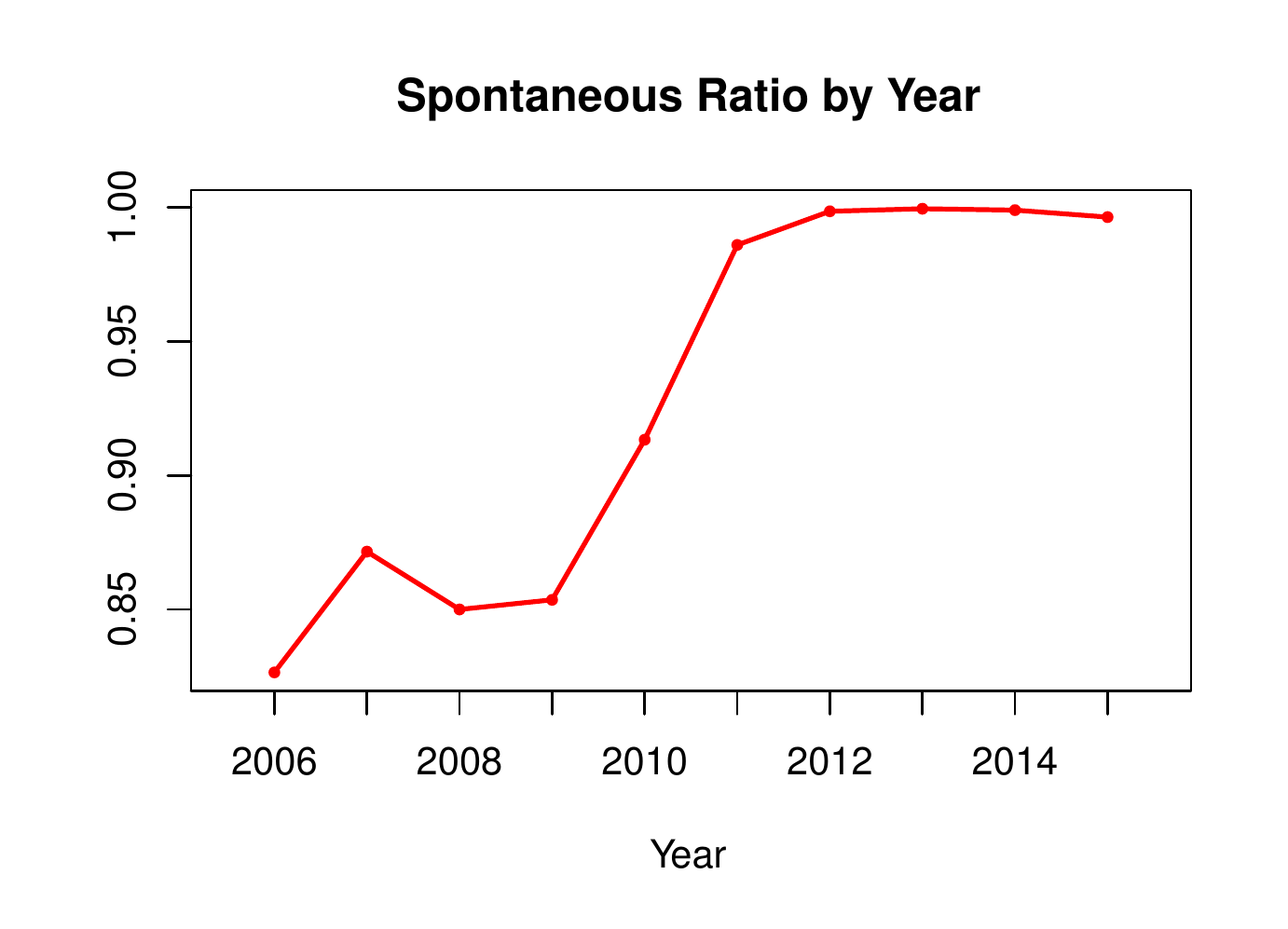}
\includegraphics[width=0.49\textwidth]{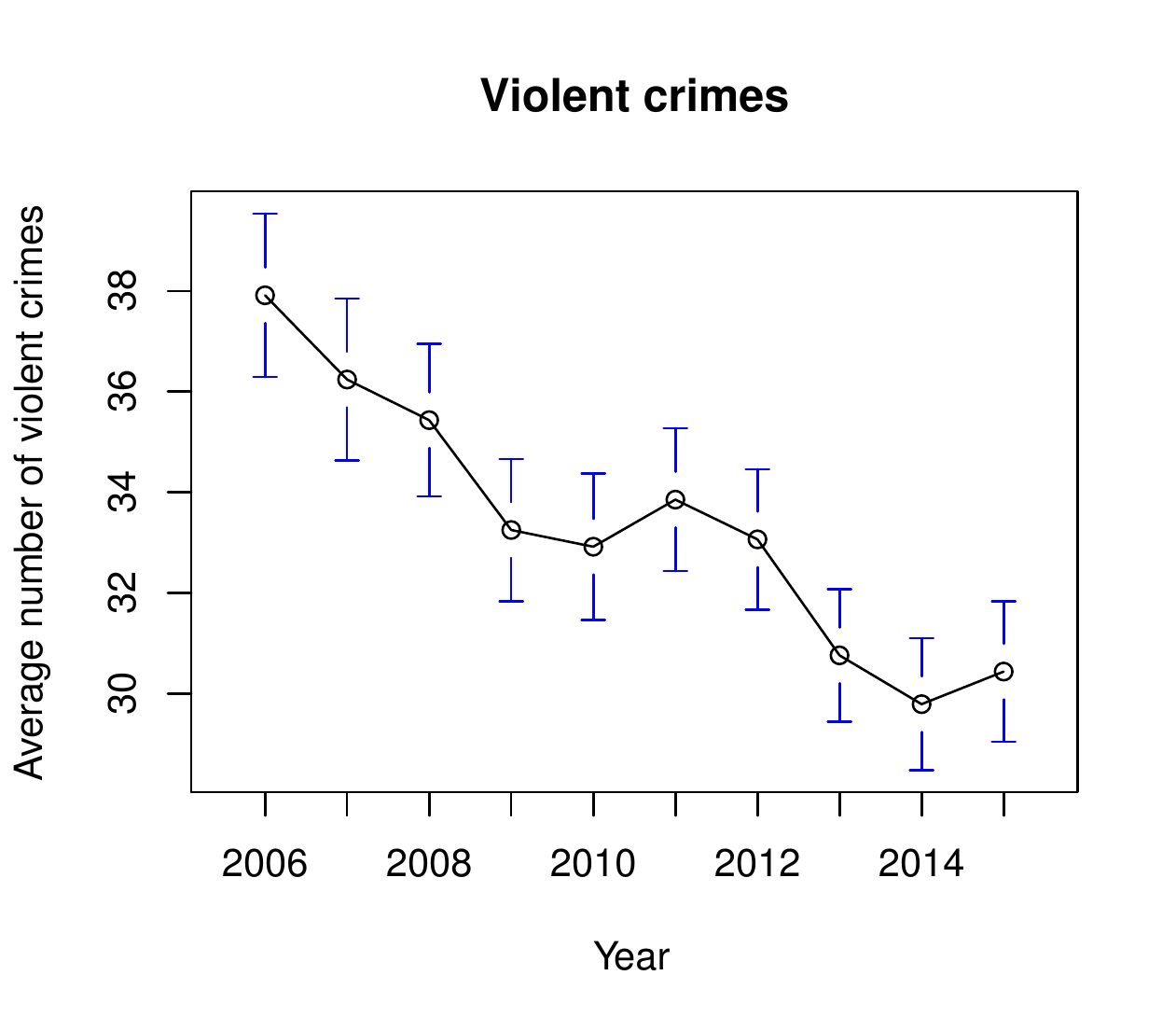}
\includegraphics[width=0.49\textwidth]{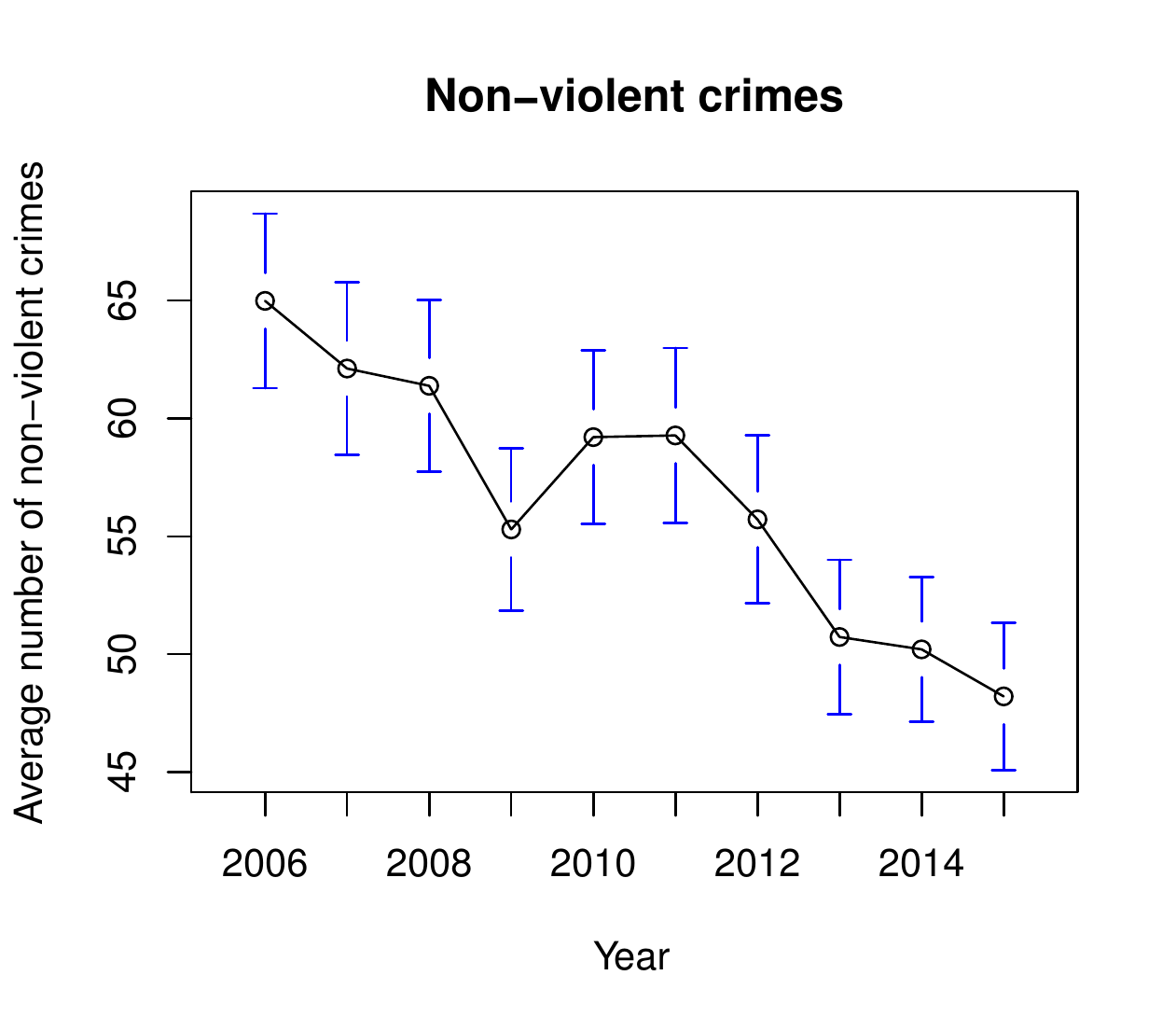}
\caption{{\bf Top row:} Yearly trends in the total number of block party events and the spontaneous proportion of block party events. {\bf Bottom row:} Yearly trends in the average number of violent and non-violent crimes per year. }
\label{timetrends}
\end{figure}

In the top row of Fig~\ref{timetrends}, we show the variation by year of the total number of block party events and the spontaneous proportion of block party events that we introduced in Section~\ref{totalnumber} and~\ref{spontaneity}.  Note that 2016 is not included in Fig~\ref{timetrends} since we only have data for part of that year.  We give monthly trends in these community vibrancy measures in Fig S2 of our supplementary materials.

We observe in the top left of Fig~\ref{timetrends} that the total number of events has been changing over time, with the number of events increasing in 2006-2008 and then decreasing from 2009 onwards.   This range of total events across Philadelphia corresponds to around 5-7 events per blockgroup per year.  In the top right of Fig~\ref{timetrends}, the spontaneous proportion is around 0.86-0.9 in the earlier years of our data but increases to 0.94 in 2010 and then very close to 1 from 2011 onwards.  In these more recent years, it seems that almost all block party permits were issued for spontaneous events rather than regular holidays.  

\section{Crime and other Neighborhood Characteristics in Philadelphia} \label{othercharacteristics}

Our crime data comes from the Philadelphia Police Department through the opendataphilly.org data portal and includes all reported crimes in Philadelphia from January 1, 2006 to December 31, 2015.   For each reported crime, we have the date, time and location in terms of GPS latitude and longitude (WGS84 decimal degrees).  Each crime is also categorized into one of several types: homicide, sex crime, armed robbery, assault, burglary, theft, motor vehicle theft, etc.

We aggregate all reported crimes within the 1,336 neighborhoods (as defined by the US Census block groups) for which we have calculated our two measures of community vibrancy.  We provide maps of the spatial distribution of crime in Fig S3 in our supplementary materials.  Since the distribution of total crimes is highly skewed across neighborhoods of Philadelphia, we will focus on the log transformation of crime in our analyses which has a more symmetric distribution.  Histograms of total crimes and the logarithm of total crimes are provided in Fig S4 in our supplementary materials.

We also make a distinction between violent, non-violent (property) crimes and vice crimes in our analysis. As defined by the Uniform Crime Reporting program of the FBI, violent crimes include homicides, rapes, robberies and aggravated assaults whereas non-violent crimes include burglaries, thefts and motor vehicle thefts. Vice crimes include drug violations, gambling, and prostitution.   

In the bottom row of Fig~\ref{timetrends}, we examine the trend over time for the two major crime categories, violent and non-violent crimes.  We see that both violent and non-violent crimes have declined over the time span of our data.   In Section~\ref{crime-aggregate}, we examine whether there is an association between our measures of community vibrancy from Section~\ref{vibrancy} and total crime incidence and then in Section~\ref{trends}, we investigate the relationship between trends over time in community vibrancy and trends over time in crime at the neighborhood level in Philadelphia.

However, before we investigate these associations between community vibrancy and crime, we first incorporate into our analysis other neighborhood characteristics that could be associated with either community vibrancy or neighborhood safety.  Specifically, we collect measures of the economic, demographic and built environment characteristics of Philadelphia neighborhoods.

Our demographic data come from the 2010 Decennial Census whereas our economic data come from the 2015 American Community Survey.  Land use data provided by the City of Philadelphia (through the opendataphilly.org data portal) gives the area and land use zoning designation for every single lot in Philadelphia.    We construct the following measures for each neighborhood (i.e. census block group) in Philadelphia:
\begin{itemize}
\item {\bf Demographic measures:} total population and the proportion of residents that identify as white, black, asian, hispanic, or other
\item {\bf Economic measures:} mean household income, poverty index (0 = poorest, 1 = wealthiest)
\item {\bf Built environment measures:} total area and the proportion of that area with designated land use of commercial, residential, vacant, transportation, industrial, park, and civic institution
\end{itemize}

%% NEW STUFF STARTS HERE (ALSO TO BE CUT AND PASTED INTO PBP)

In Table~\ref{table:data}, we provide additional details for each data source as well as the raw data variables used to construct the measures above.  Similar measures have been used to capture surrounding neighborhood context in other studies of the association between the built environment and crime.  \cite{HumJenSma20} use US Census data for Philadelphia to create demographic measures such as population count and racial proportions, as well as economic such as per capita household income.  Additional details about the poverty index that we use in this paper are given in \cite{HumJenSma20}.  Race and poverty measures were also used by \cite{BraCheMac11} as control variables in their evaluation of the effects of vacant lot greening on crime and health outcomes.

\cite{Mac15} reviews empirical research on associations between crime and quantitative measures of the built environment, including proportions of commercial, residential and mixed land use.  Land use characteristics such as presence of commercial or industrial property, vacant lots or single vs. multi-family residential units were used to evaluate crime incidence around bus stops \citep{LouLigIse01} and green line transit stations \citep{LouLigIse02} in Los Angeles.  \cite{Zha16} used measures based on the share of residential vs. commercial land use in order to investigate the association between land use, crime and public transit ridership.  \cite{HumJenSma20} also use land use zoning data in Philadelphia to create measures such as the proportion of vacant land and the proportion of commercial vs. residential land use. 

\begin{table}[!h]
\caption{Additional details about each data source used in our analysis.  For each data type, we list the raw data variables used and any measures constructed from those data.}
\centering
\includegraphics[width=\textwidth]{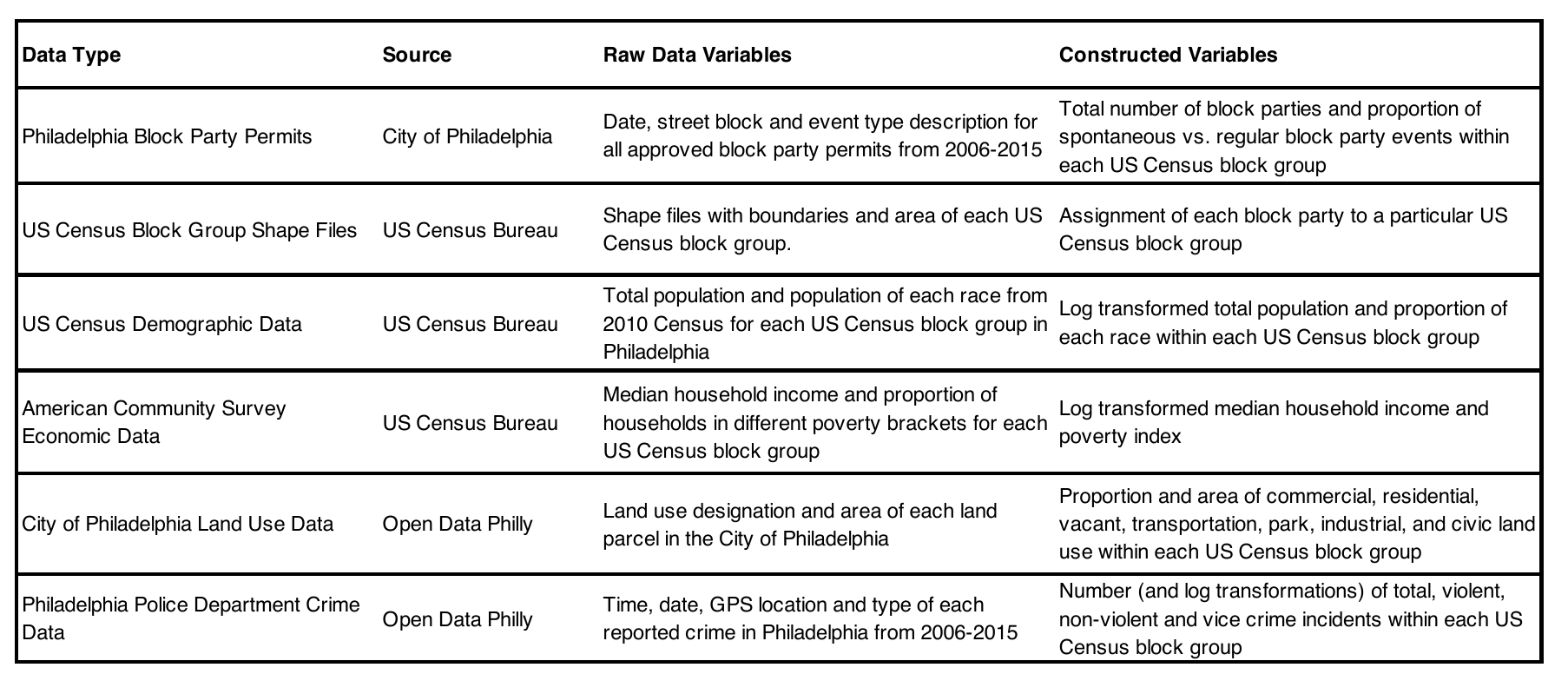}
\label{table:data}
\end{table}

%% NEW STUFF ENDS HERE (ALSO TO BE CUT AND PASTED INTO PBP)

In Fig S5 of our supplementary materials, we provides correlations between these demographic, economic and built environment measures and our measures of community vibrancy and crime.    We observe that spontaneous proportion of block parties is not strongly correlated with any of these other neighborhood characteristics.  However, the total number of block party permits is correlated with both economic measures (median income and poverty index) as well as the proportion of black residents in a neighborhood.  We also see that crime is strongly correlated with several other neighborhood characteristics.

\begin{figure}[!h]
\centering
\includegraphics[width=0.32\textwidth]{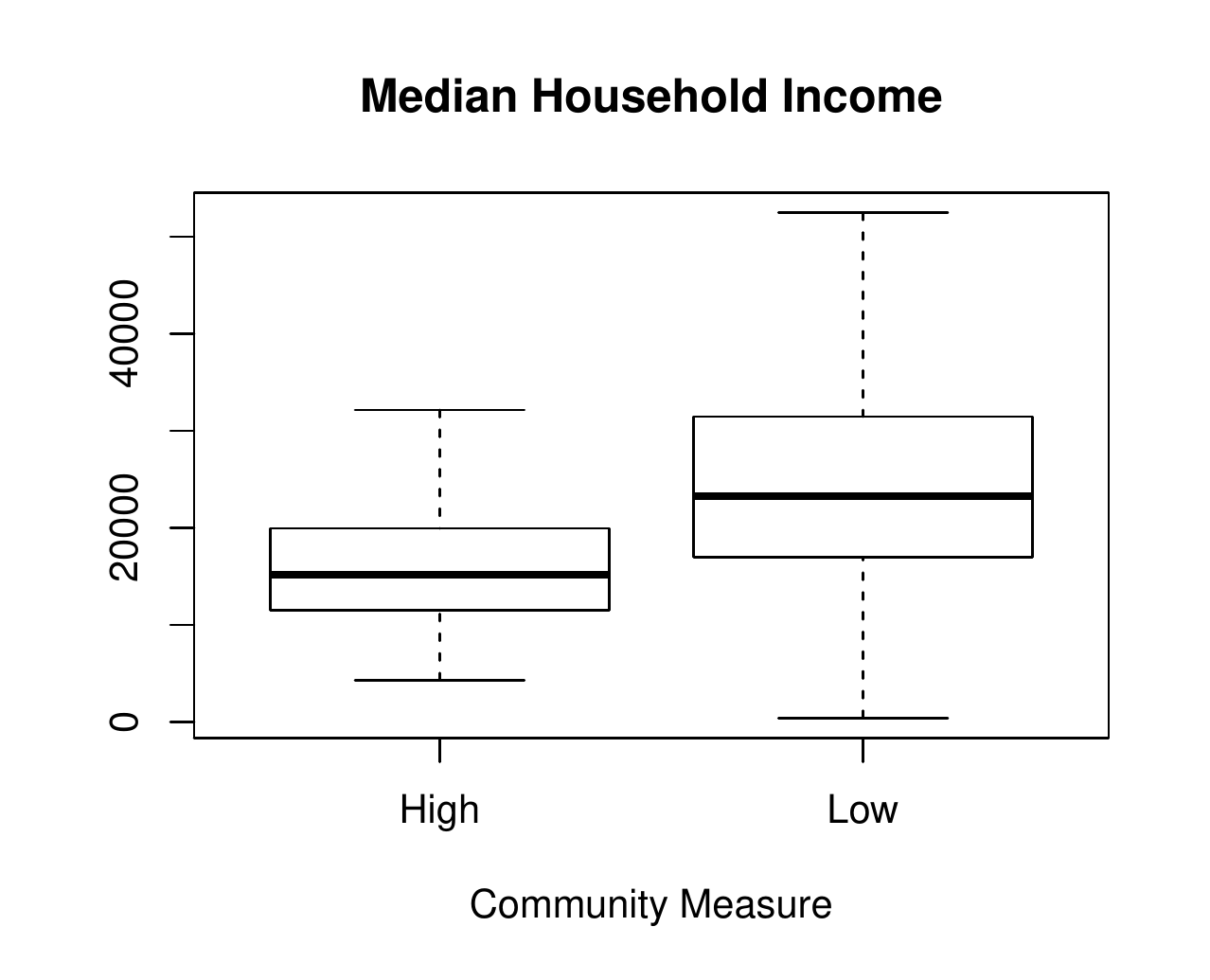}
\includegraphics[width=0.32\textwidth]{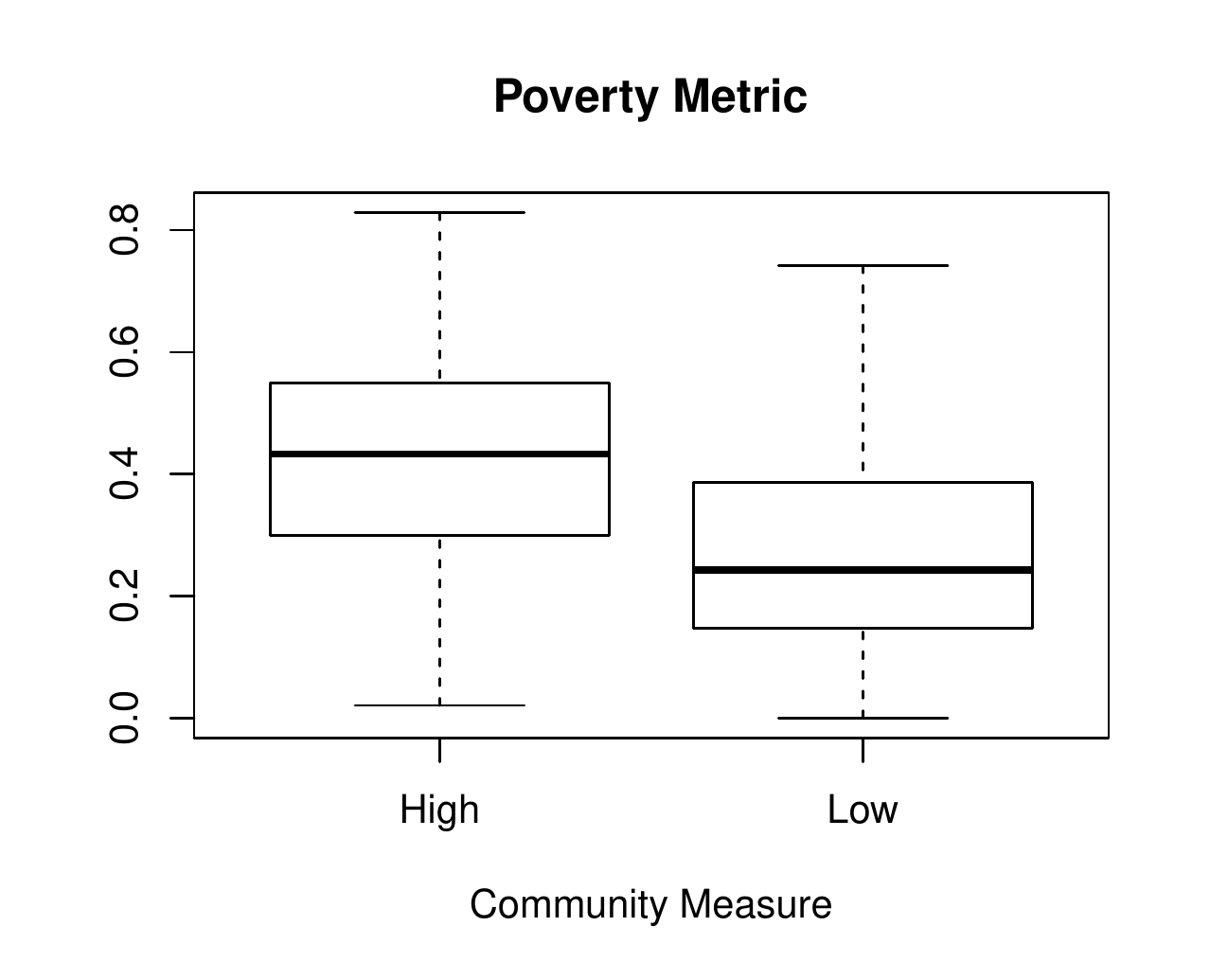}
\includegraphics[width=0.32\textwidth]{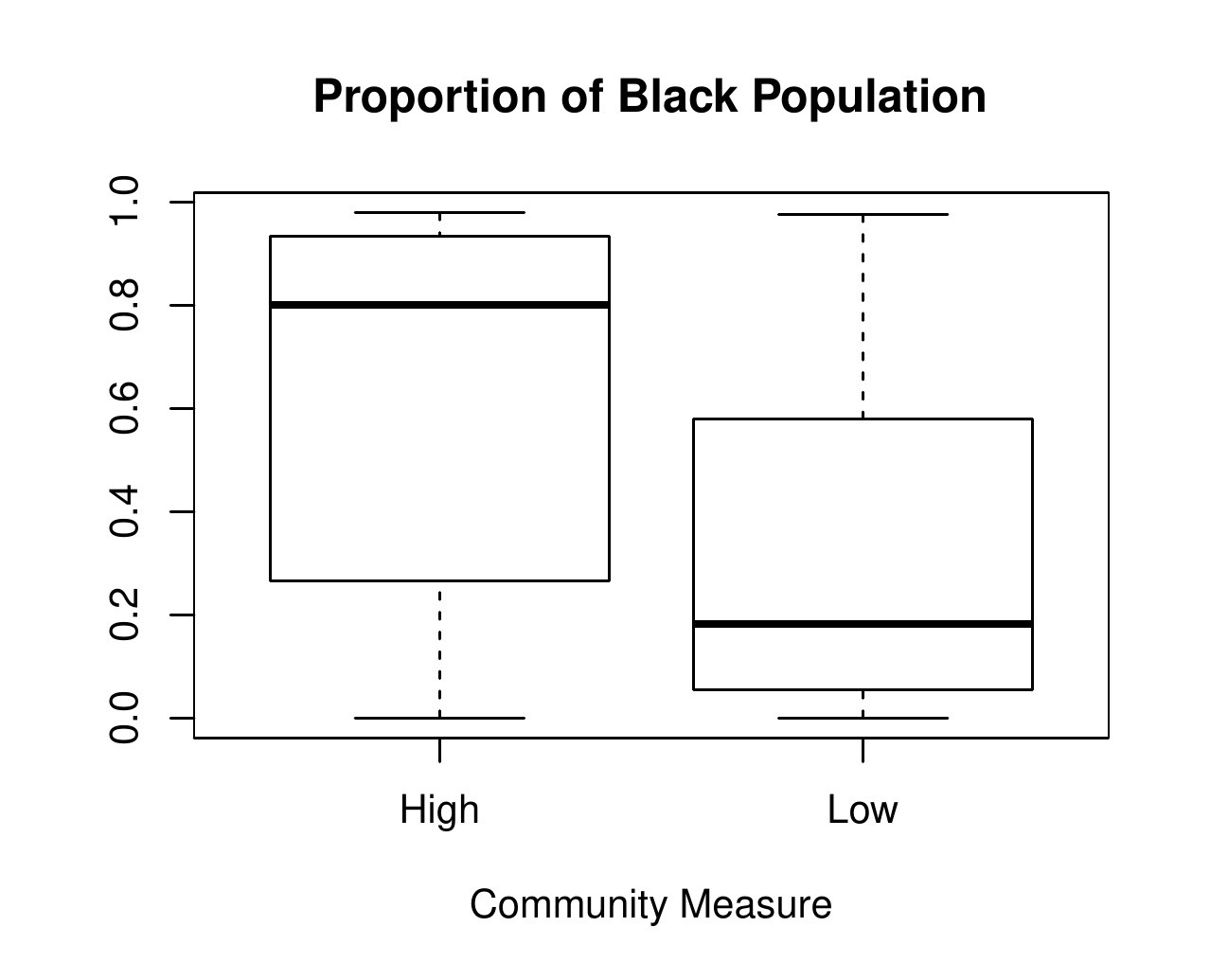}
\caption{Distribution of median household income, poverty metric, and proportion of Black population between high and low vibrancy neighborhoods in Philadelphia.}
\label{fig:boxplots}
\end{figure}

The association between community vibrancy, crime incidence and these other neighborhood characteristics means that any comparison of crime incidence that we make between high vibrancy and low vibrancy neighborhoods could be confounded by an imbalance on these other neighborhood characteristics.   This imbalance is apparent in Fig~\ref{fig:boxplots} where we see significant differences in median household income, poverty metric, and proportion of Black population between high and low vibrancy neighborhoods in Philadelphia.

In our investigation into the relationship between community vibrancy and safety, we will employ two different approaches to account for imbalance in these other neighborhood characteristics: linear regression modeling and propensity score matching.

\section{Association between Overall Community Vibrancy and Crime}\label{crime-aggregate}

In this section, we investigate the relationship between crime incidence and our measures of community vibrancy at the neighborhood level over the entire 2006-2016 time span of our crime and block party permit data.  We will employ two different analyses in order to account for other characteristics of Philadelphia neighborhoods: regression modeling and propensity score matching.  

\subsection{Linear Regression Analysis of Total Crime and Community Vibrancy}\label{aggregate.regression}

In this regression approach, we consider total crime incidence from 2006-2016 within each neighborhood as our outcome variable and we are interested in whether our measures of community vibrancy are significant predictors of this outcome while controlling for other neighborhood characteristics.  

Specifically, we consider the following linear model for the logarithm of total crime incidence $y_{i}$ in block group $i$:  
\begin{eqnarray}
\log \left( y_{i} \right) = \beta_0 + \bbeta \cdot \X_{i} + \phi \cdot C_{i} + \epsilon_{i} \label{regressioneqn}
\end{eqnarray}
where $\X_i$ are the demographic, economic, and land use characteristics of neighborhood $i$ as outlined in Section~\ref{othercharacteristics} and $C_i$ is one of our community vibrancy measures, either the number of total block party permits or the spontaneous proportion for neighborhood $i$.     We are specifically interested in whether the coefficient $\phi$ is non-zero, which would imply that particular measure of community vibrancy $ C_{i}$ is predictive of total crime incidence beyond the other neighborhood characteristics included in the model. 

We use a log transformation of total crime incidence $y_i$ since Fig S4 in our supplementary materials suggests that the log scale for crime is a more reasonable fit to the assumption of normally distributed errors $\epsilon_{i}$.   However, we also consider an alternative regression approach where the total crime incidence $y_i$ is directly modeled as a negative binomial random variable that is a linear function of the same predictor variables as in eqn (\ref{regressioneqn}).

We will compare the results from four different regressions that represent each combination of our two community vibrancy measures and our two regression model specifications, 
\begin{enumerate}
\item Ordinary least squares (OLS) regression of the logarithm of total crime incidence $\log (y_{i})$ on the number of total events $C_i$ and other neighborhood characteristics $\X_i$
\item Ordinary least squares (OLS) regression of the logarithm of total crime incidence $\log (y_{i})$ on the spontaneous proportion $C_i$ and other neighborhood characteristics $\X_i$
\item Negative binomial regression of total crime incidence $\log (y_{i})$ on the number of total events $C_i$ and other neighborhood characteristics $\X_i$
\item Negative binomial regression of total crime incidence $\log (y_{i})$ on the spontaneous proportion $C_i$ and other neighborhood characteristics $\X_i$
\end{enumerate}

As detailed in Section~\ref{othercharacteristics}, our set of other neighborhood characteristics $\X_i$ for each block group $i$ consist of the total population and fraction of white, black, asian and hispanic residents, our poverty metric and the log of mean household income, and the total area and fraction of that area that is zoned as vacant, commercial or residential.  Table S1 in our supplementary materials displays the parameter estimates and model fit statistics for the four regression models outlined above.  The OLS regression models are a better fit to the data than the negative binomial regression models in terms of root mean square error (RMSE).  

We see in Table S1 that most neighborhood characteristics have significant partial effects, which suggests that each of these economic, demographic and land use characteristics have an association with crime, even after accounting for the other characteristics included in the model.  Higher levels of poverty and larger commercial proportions are associated with higher levels of total crime in each of the four models, whereas higher proportions of park space and residential land use are associated with lower levels of total crime. 

However, our primary interest is the association between our measures of community vibrancy and crime, having controlled for these other neighborhood characteristics.   In Table S1, we see that the number of total permits is significantly positively associated with total crimes (models 1 and 3), whereas the spontaneous proportion is non-significantly negatively associated with total crimes (models 2 and 4).  

In particular, we see that a one unit increase in the number of block party permits is associated with a 0.2\% increase in the number of total crimes, holding all other variables constant (from model 1).  We also see that a 10\% increase in the spontaneous proportion is associated with a 2.8\% decrease in the number of total crimes, though this is association is not statistically significant (from model 2).   

We found highly similar results when we ran regression models with (a) just violent crimes, (b) just non-violent crimes or (c) just vice crimes as outcome variables.   Tables and details for these additional regression models are also given in our supplementary materials.  

It is interesting to see that our two measures of community vibrancy have very different associations with crime.   Greater numbers of total permits is associated with a greater number of total crimes whereas a larger spontaneity proportion is associated with fewer total crimes.  The opposing directions of these associations suggest that our two measures are capturing quite different aspects of community and the relationship between community and crime.  

To the extent that spontaneous block party events are more indicative of concentrated community cohesion among a few households, the association between larger spontaneous proportion and fewer crimes suggests that this localized cohesion may signal greater guardianship than the overall number of block parties in a community.  It is also possible that spontaneous block party events are more inclusive (compared to say, religious events) to newer residents which could also increase the collective efficacy towards crime prevention within a community.

As an alternative approach to evaluating the relationship between our two measures of community vibrancy and crime incidence, we employ a propensity score matching analysis in Section~\ref{sec:results:overall:propensity} below.

\subsection{Propensity Score Matching Analysis of Total Crime and Community Vibrancy} \label{sec:results:overall:propensity}

In Section~\ref{aggregate.regression}, we used regression models to estimate the association between community vibrancy and total crime, while accounting for the demographic, economic and land use characteristics of Philadelphia neighborhoods.  Matching analyses are an alternative approach for isolating the relationship between community vibrancy and total crime from these other neighborhood characteristics. 

In this approach, we create artificial experiments consisting of matched pairs of neighborhoods that have highly similar demographic and economic characteristics but differ substantially in terms of their measures of community vibrancy, which allows us to isolate the relationship between community vibrancy and crime.

We set up two different experiments to investigate each of our two measures of community vibrancy.  In the first experiment, we categorize all Philadelphia neighborhoods into a ``treatment" group vs. ``control" group based on whether their total number of block party permits were above or below the city-wide median of 42.5 block parties.  In the second experiment, we categorize all Philadelphia neighborhoods into a ``treatment" group vs. ``control" group based on whether their spontaneity proportion was above or below the city-wide median of 0.962. 

Within each experiment, our goal is to create pairs of neighborhoods consisting of one treatment neighborhood and one control neighborhood that both share highly similar economic, demographic and land use characteristics.  These matched pairs allow us to evaluate the association between crime and our two community vibrancy measures based on within-pair comparisons that are balanced on these other neighborhood characteristics.

We create these matched pairs using a propensity score matching procedure \citep{RosRub83}.  The {\it propensity score} for each unit (neighborhood) in our analysis is the estimated probability that a particular unit (neighborhood) receives the treatment (high community vibrancy) based on other neighborhood characteristics.   We estimate these propensity scores using a logistic regression model with the treatment vs. control indicator as the outcome and the demographic, economic and land use measures for each neighborhood as predictors.  

Two neighborhoods with highly similar demographic, economic and land use characteristics will have highly similar propensity scores.  For each neighborhood in the treatment group (e.g. having a large number of block parties), we will create a matched pair by finding a neighborhood in the control group (e.g. having a small number of block parties) that has a highly similar propensity score.  Thus, within each matched pair we have an ``apples-to-apples" comparison of two neighborhoods that have differ in terms of high vs. low community vibrancy but have highly similar other neighborhood characteristics.

\begin{figure}[!ht]
\centering
\includegraphics[width=0.47\textwidth]{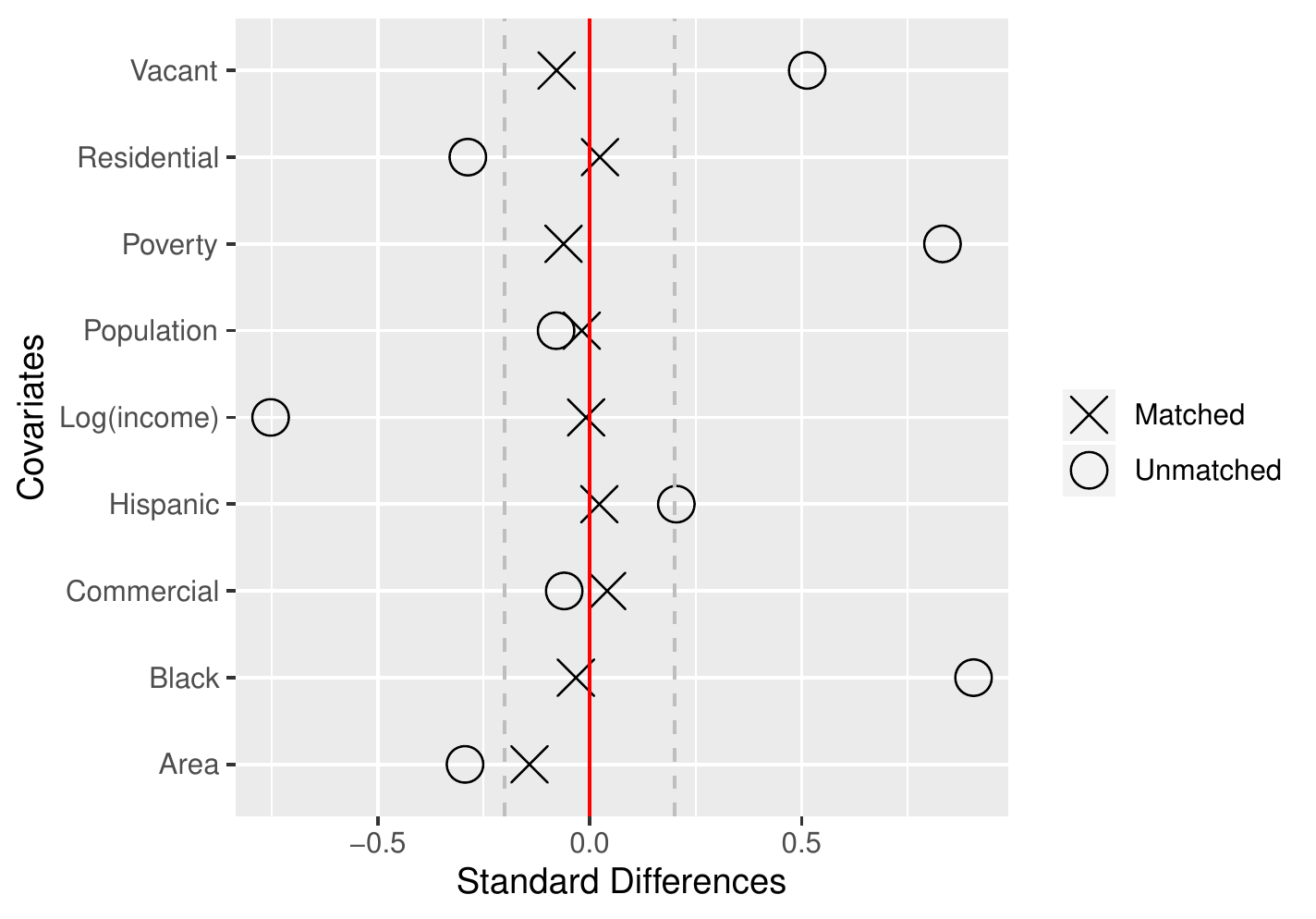}
\includegraphics[width=0.47\textwidth]{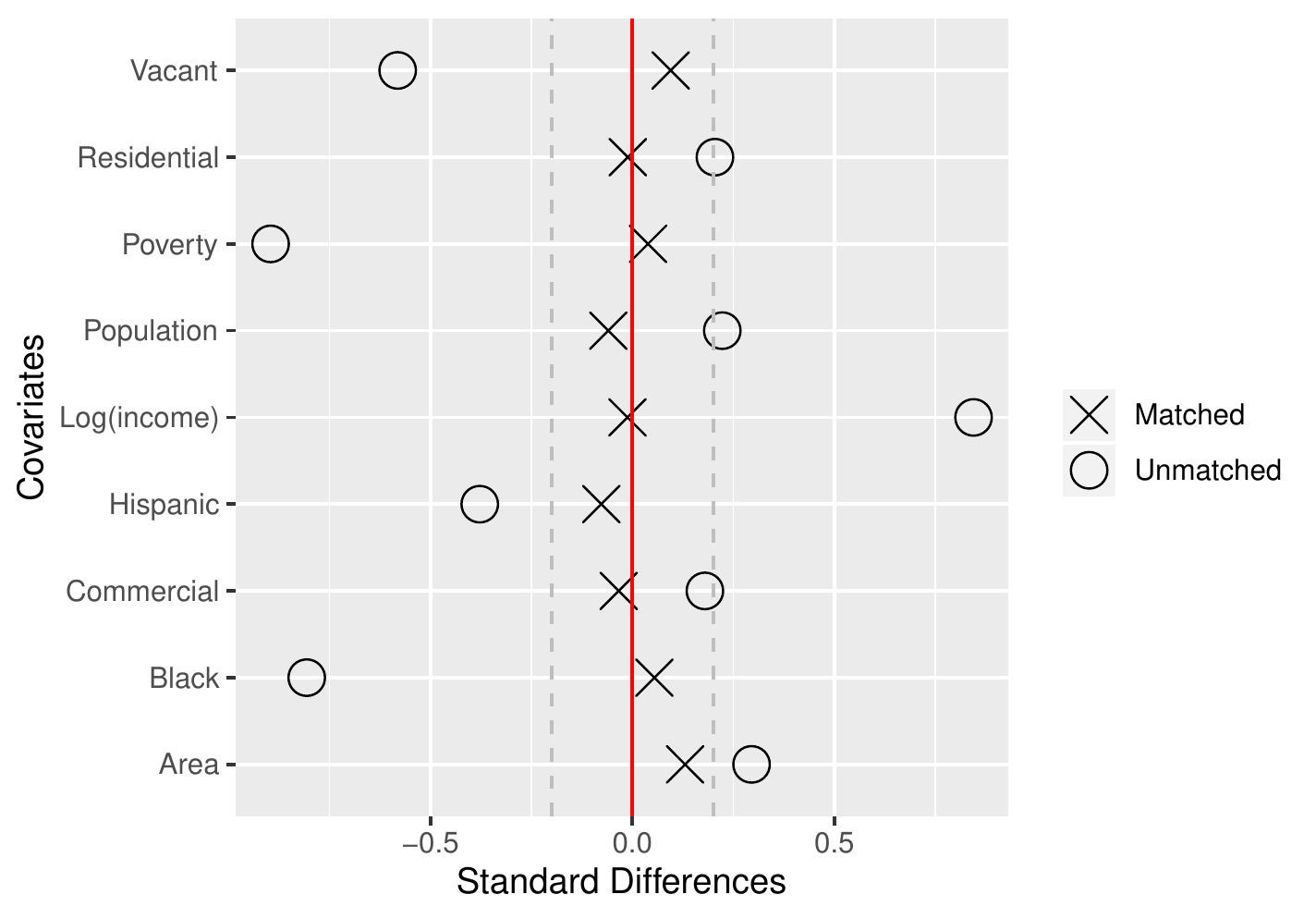} \\
\includegraphics[width=0.49\textwidth]{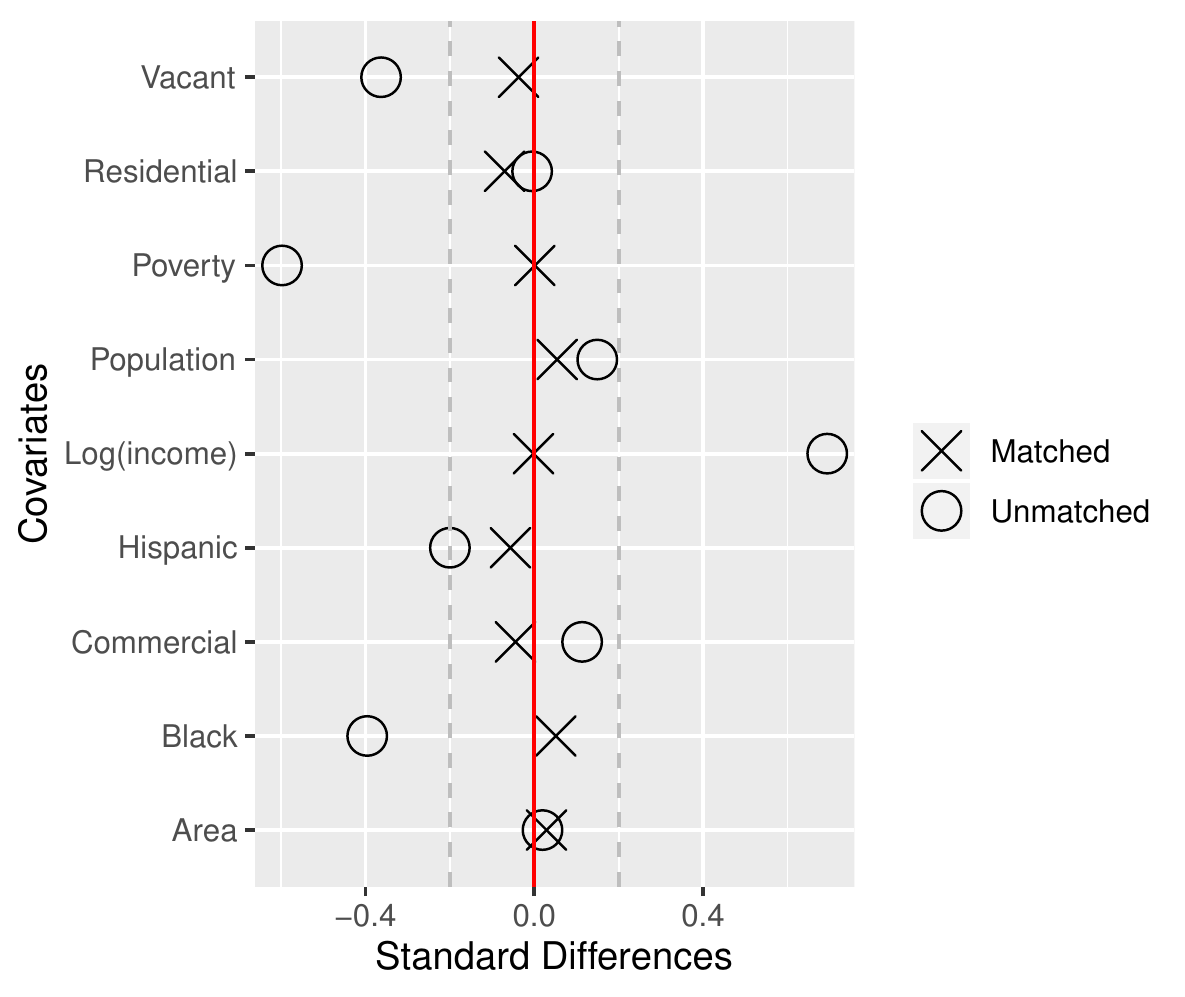}
\includegraphics[width=0.49\textwidth]{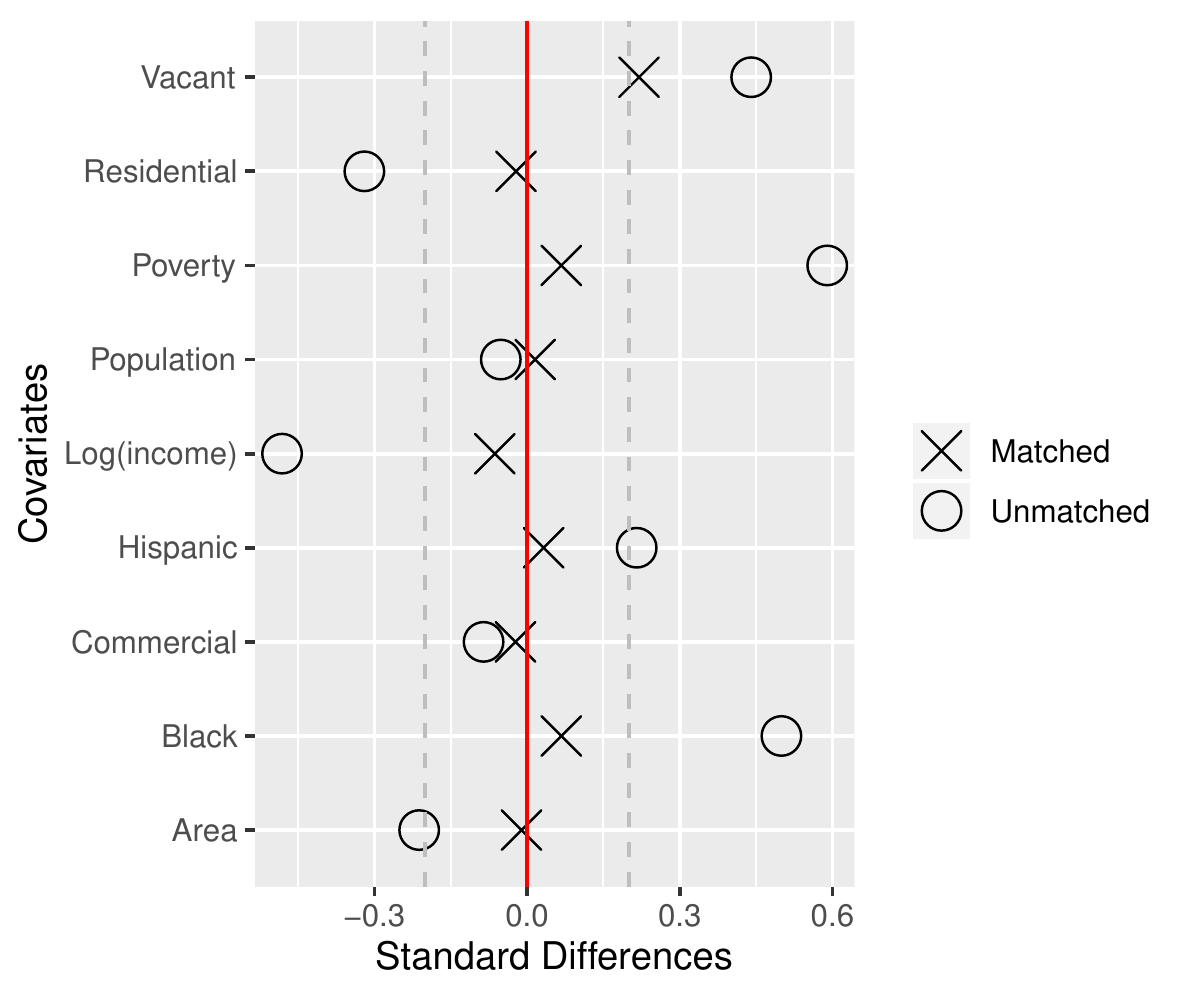}
\caption{{\bf Top row} is the standardized differences between neighborhoods with high vs. low community vibrancy, both before and after propensity score matching. {\bf Top Left:} total number of permits as the measure used to define the high vs. low community vibrancy group. {\bf Top Right}: spontaneous proportion as the high vs. low community vibrancy measure.  {\bf Bottow row} is the standardized differences between neighborhoods with increasing trends over time in community vibrancy or not.  {\bf Bottom Left:} treatment group is neighborhoods that have a significantly increasing trend over time in block party permits.  {\bf Bottom Right:} treatment group is neighborhoods that have a significantly increasing trend over time in spontaneous proportion.}
\label{love1}
\end{figure}

In the top row of Fig~\ref{love1}, we evaluate the balance in other neighborhood characteristics that we have achieved with our propensity score matching procedure.   Specifically, we compare the standardized differences in each neighborhood characteristic between high vs. low community vibrancy neighborhoods before matching to the standardized differences within our matched pairs.  We give separate plots for our two different experiments where either the total number of block parties (top left) or the spontaneous proportion (top right) were used to define our high vs. low community vibrancy groups. 

We see in Fig~\ref{love1} that our propensity score matching procedure has created matched pairs of neighborhoods with almost no difference in their demographic, economic and land use characteristics.  This balance in other neighborhood characteristics enables us to better isolate the relationship between our two measures of community vibrancy and total crime.  We then use our created sets of matched pairs to estimate the effect of having high community vibrancy on total crime at the neighborhood level.  

When using the total number of block party permits as our measure of community vibrancy, we find that the average within-pair difference in log total crimes is 0.223 between the high vibrancy neighborhood and the low vibrancy neighborhood and the 95\% confidence interval on that average within-pair difference is [0.173, 0.273].  This interval suggests that neighborhoods with a high number of block party permits have roughly between 1.2-1.3 times as many total crimes as neighborhoods with a low number of block party permits.  So we find that total crimes are significantly higher in neighborhoods with a large number of block party permits compared to their matching neighborhoods that have a small number of block party permits.    

When using the spontaneous proportion as our measure of community vibrancy, we find that the average within-pair difference in log total crimes is -0.991 between the high spontaneous proportion neighborhood and the low spontaneous proportion neighborhood.  The 95\% confidence interval on that average within-pair difference is [-0.148, -0.050].  This interval suggests that neighborhoods with a high spontaneous proportion have roughly between 0.86-0.95 times as many total crimes as neighborhoods with a low spontaneous proportion.   So we find that total crimes are significantly lower in neighborhoods with a high spontaneous proportion compared to their matching neighborhoods that have a low spontaneous proportion.
    
This propensity score matching analyses indicate that our two measures of community vibrancy have significant associations with total crime over the 2006-2016 time period of our data.  These results confirm our earlier regression analyses that these associations are in opposing directions for our two measures of community vibrancy: greater numbers of total permits are associated with a greater number of total crimes and a greater spontaneous proportion is associated with fewer total crimes.  

In the following section, we evaluate how these measures of community vibrancy and crime have changed together over time.  

\section{Trends in Block Parties and Crime over Time}\label{trends}

In Section~\ref{crime-aggregate}, we found significant associations between overall levels of crime and community vibrancy at the neighborhood level, when accounting for other characteristics of those neighborhoods.  However, levels of crime and our measures of community vibrancy have all changed substantially over this time period across Philadelphia.  In this section, we investigate the relationship between changes in crime incidence over time and the changes in community vibrancy over time at the neighborhood level.   

As a reminder, we can compare the overall trends in yearly crime incidence to the trends by year in our two community vibrancy measures in Fig~\ref{timetrends}.  We see that both the number of permits and total crime incidence have a decreasing trend while the spontaneity proportion has an increasing trend over the time span of our data. 

However, trends over time in either crime incidence or community vibrancy can vary substantially between different neighborhoods across the city.  We are interested in the association between trends over time in crime incidence and trends over time in community vibrancy across these different neighborhoods.   We will again employ two different analyses in order to account for other characteristics of Philadelphia neighborhoods: regression modeling and propensity score matching.

\subsection{Regression Analysis of Trends over Time}

We summarize the trend over time in crime within each neighborhood by fitting a separate linear regression of the yearly number of total crimes within each neighborhood on year, and then classifying neighborhoods according to their slope on crime over time.   Only 18 neighborhoods (1.4\%) had a significantly positive linear trend in crime over time, whereas 540 neighborhoods (42.4\%) had a significantly negative linear trend in crime over time.

Similarly, we summarize the trend over time in community vibrancy within each neighborhood by fitting a separate linear regression of the yearly number of block party permits within each neighborhood on year, and then classifying neighborhoods according to their slope on number of permits over time.   Only 94 neighborhoods (7.4\%) had a significantly positive linear trend in number of permits over time, whereas 184 neighborhoods (14.4\%) had a significantly negative linear trend in number of permits over time.

We will focus our regression analyses on determining the neighborhoods factors that are predictive of whether or not a neighborhood has a significant trend over time in either crime or our measures of community vibrancy.   Specifically, we fit the four different logistic regression models enumerated below: 
\begin{enumerate}
\item Logistic regression with significantly increasing trend in community (or not) as the outcome and neighborhood characteristics $\X_i$ (including indicators of trends in crime) as the predictors 
\item Logistic regression with significantly decreasing trend in community (or not) as the outcome and neighborhood characteristics $\X_i$ (including indicators of trends in crime) as the predictors
\item Logistic regression with significantly increasing trend in crime (or not) as the outcome and neighborhood characteristics $\X_i$ (including indicators of trends in community) as the predictors
\item Logistic regression with significantly decreasing trend in crime (or not) as the outcome and neighborhood characteristics $\X_i$ (including indicators of trends in community) as the predictors
\end{enumerate}

Table S2 in our supplementary materials displays the parameter estimates and model fit statistics for the four logistic regression models listed above, where we use the number of block party permits as our measure of community.   We see in Table S2 that log income is a strong predictor of significantly increasing trends in block party permits and that vacant proportion is a strong predictor of significantly decreasing trends in block party permits.  We also see that industrial land use is a strong predictor of a significantly increasing trend in crime and that the Hispanic proportion is a strong predictor of a significantly decreasing trend in crime.    

In Table S2, we see that trends in crimes are not predictive of trends in the number of block party permits and vice versa. However, there are so few neighborhoods with significantly increasing trends in either block party permits or crimes, which gives us limited power to detect subtle associations.   

We fit the same four logistic regression models but using spontaneous proportion as our measure of community and the results are given in Table S3 of our supplementary materials.  In Table S3, we see that trends in crimes are also not predictive of trends in the spontaneous proportion and vice versa.  These results suggest that there are no strong associations between trends over time in crime and trends over time in our two measures of community vibrancy.  

We further investigate these longitudinal trends with an alternative analysis based on propensity score matching in Section~\ref{matched-timetrend}.

\subsection{Propensity Score Matching for Examining Trends over Time}\label{matched-timetrend}

Similar to our approach in Section~\ref{sec:results:overall:propensity}, we create artificial experiments consisting of matched pairs of neighborhoods that have highly similar demographic and economic characteristics but the two neighborhoods within each pair differ substantially in terms of their trend over time in community vibrancy.  This approach allows us to isolate the relationship between trends over time in community vibrancy and trends over time in crime.

For example, we can categorize all neighborhoods based on whether they have a significantly positive trend in the number of block party permits or not.  We label neighborhoods with a significantly positive trend in the number of block party permits as the ``treatment" group and label all other neighborhoods as the ``control" group.  Just as in Section~\ref{sec:results:overall:propensity}, we fit a logistic regression with these treatment vs. control labels as the outcome variable and all other neighborhood factors (demographic, economic and land use) as predictor variables of that outcome.  From this fitted model, the probability of a neighborhood being in the treatment group is called the {\it propensity score} for that neighborhood.  

We then match up each neighborhood in the treatment group with a neighborhood from the control group with the closest possible propensity score.  In this way, we form a set of matched pairs where each pair of neighborhoods have highly similar demographic, economic and land use characteristics but one of those neighborhoods has a significantly positive trend in the number of block party permits and the other neighborhood does not.

The bottow row of Fig~\ref{love1} compares the standardized differences between neighborhoods before and after propensity score matching for two of the experiments that we perform.  In the bottom left plot, the treatment group is neighborhoods that have a significantly increasing trend over time in block party permits whereas in the bottom right plot, the treatment group is neighborhoods that have a significantly increasing trend over time in spontaneous proportion.   We see that, for both experiments, our matching procedure has created pairs of neighborhoods with almost no difference in their demographic, economic and land use characteristics, which makes for a more balanced comparison of crime between neighborhoods that have significantly positive trends over time in either of our two community vibrancy measures. 

We considered twelve different propensity score matching experiments with each experiment being a different combination of  four definitions for the treatment variable and three crime outcomes.  The four treatment variables considered were: 1. having a significantly positive trend over time in block party permits, 2. having a significantly negative trend over time in block party permits, 3. having a significantly positive trend over time in the spontaneous proportion, and 4. having a significantly negative trend over time in spontaneous proportion.   For each of these different treatments, we evaluated our matched pairs of neighborhoods for differences in three crime trend outcomes: 1. the slope on the trend over time in total crime, 2. an indicator for a significantly positive crime trend (or not) and 3. an indicator for a significantly negative crime trend (or not).   

Table~\ref{matchedpairs} gives the average within-pair differences between the treatment and control groups (and 95\% confidence intervals for those averages) for all twelve combinations outlined above.  We see that 11 of the 12 comparisons do not yield statistically significant results.   However, we do find that neighborhoods which have a significantly positive trend in their spontaneous proportion also show significantly negative trends over time in total crimes.  This is the only significant association we have been able to detect between trends over time in crime and trends over time in our two measures of community vibrancy.  

\begin{table}[!htpb]
\centering
\small
\caption{Average within-pair differences between the treatment and control groups (and 95\% confidence intervals) for all twelve combinations of four treatment variables (columns) and crime outcomes (rows).  For the ``crime slope" outcome, the difference between slopes is provided, whereas for the ``Crime $+$" and "Crime $-$" indicators, the odds ratio is provided}
\label{matchedpairs}
\begin{tabular}{l|cccc}
\\[-1.8ex]\hline 
\hline \\[-1.8ex] 
 & \multicolumn{4}{c}{Treatment}                           \\ 
\hline \\[-1.8ex] 
Outcome &
  \# Permits $+$ &
  \# Permits $-$ &
   Spont $+$ &
  Spont $-$ \\ 
\hline \\[-1.8ex] 
Crime slope &
  \begin{tabular}[c]{@{}c@{}}$-1.4341$\\ $[-4.7230, 1.8548]$\end{tabular} &
  \begin{tabular}[c]{@{}c@{}}$-0.6030671$\\ $[-2.9085, 1.7024]$\end{tabular} &
   \begin{tabular}[c]{@{}c@{}}$-2.1928^{***}$\\ $[-3.8507, -0.5350]$\end{tabular} &
  \begin{tabular}[c]{@{}c@{}}$0.6705$\\ $[-1.1154, 2.4563]$\end{tabular} \\
Crime $+$ &
  \begin{tabular}[c]{@{}c@{}}$1.0085$\\ $[0.9356, 1.0509]$\end{tabular} &
  \begin{tabular}[c]{@{}c@{}}$ 0.9933$\\ $[0.9588, 1.0290]$\end{tabular}  &
   \begin{tabular}[c]{@{}c@{}}$1.0053$\\ $[0.9724, 1.0393]$\end{tabular} &
  \begin{tabular}[c]{@{}c@{}}$0.9902^{\dagger}$\\ $[0.9367, 1.0468]$\end{tabular} \\
%  \begin{tabular}[c]{@{}c@{}}$0.0053$\\ $[-0.0280, 0.0386]$\end{tabular} &
%  \begin{tabular}[c]{@{}c@{}}$-0.0098$\\ $[-0.0654, 0.0458]$\end{tabular} &
%  \begin{tabular}[c]{@{}c@{}}$-0.0084$\\ $[-0.0666, 0.0497]$\end{tabular} &
%  \begin{tabular}[c]{@{}c@{}}$-0.0067$\\ $[-0.0421,  0.0286]$\end{tabular} \\
Crime $-$ &
  \begin{tabular}[c]{@{}c@{}}$1.0393$\\ $[0.8852, 1.2201]$\end{tabular} &
  \begin{tabular}[c]{@{}c@{}}$1.0503$\\ $[0.9364, 1.1781]$\end{tabular} &
    \begin{tabular}[c]{@{}c@{}}$ 1.0558$\\ $[0.9715,1.1475]$\end{tabular} &
  \begin{tabular}[c]{@{}c@{}}$1.1560$\\ $[0.8970, 1.4899]$\end{tabular} \\  \\[-1.8ex] 
%  \begin{tabular}[c]{@{}c@{}}$0.0543$\\ $[-0.0289, 0.1376]$\end{tabular} &
%  \begin{tabular}[c]{@{}c@{}}$0.1450$\\ $[-0.1087, 0.3987]$\end{tabular} &
%  \begin{tabular}[c]{@{}c@{}}$0.0385$\\ $[-0.1219, 0.1989]$\end{tabular} &
%  \begin{tabular}[c]{@{}c@{}}$0.0491$\\ $[-0.0657, 0.1639]$\end{tabular} \\  \\[-1.8ex] 
\hline 
\hline \\[-1.8ex] 
\textit{Note:}  & \multicolumn{4}{l}{\scriptsize $^{***}$p$<$0.05; $^{**}$p$<$0.01; $^{***}$p$<$0.001 with Wilcoxon} \\ [-0.8ex] 
 & \multicolumn{4}{l}{\scriptsize $^{\dagger}$Estimates from many to one matching rather than 1:1 due to imbalance}
\end{tabular}
\end{table}

\section{Summary and Discussion}\label{discussion}

In this paper, we explore the relationship between crime incidence at the neighborhood level and two measures of community vibrancy created from a unique dataset of block party permit approvals in the city of Philadelphia.   As outlined in Sections~\ref{introduction} and \ref{vibrancy}, we design these two measures to capture potentially different aspects of community with our first measure reflecting the overall volume of block party events whereas our second measure reflects the distinction between regular versus spontaneous block party events as these different types could be associated with different levels of community engagement.  

In order to properly analyze the relationship between our measures of community vibrancy and crime, we must account for the economic, demographic and land use characteristics of these neighborhoods which may also have an influence on both community vibrancy and crime incidence.   We employ two statistical techniques, regression modeling and propensity score matching, in order to isolate the association between crime and community vibrancy while controlling for other neighborhood characteristics.  

We find significant associations between aggregate levels of crime and our two measures of community vibrancy at the neighborhood level, while accounting for other characteristics of those neighborhoods.   Neighborhoods with more block parties have a significantly higher crime rate, while those holding a greater proportion of spontaneous events have a significantly lower crime rate.  We also find that neighborhoods which have a significantly positive trend in their spontaneous proportion also show significantly negative trends in total crimes over time.

Previous studies suggest that public signals of community cohesion, guardianship and collective efficacy can lead to crime prevention \citep{SamGro89, SamRauEar97}.  The different associations with crime incidence we find for our two measures of community vibrancy may indicate that different aspects of community cohesion are being captured by the total volume of block party events versus the type of block party events.   In particular, we see that a greater number of block parties is associated with increased crime but that a larger spontaneous proportion is associated with fewer total crimes.  Spontaneous block parties may indicate more concentrated cohesion among a few households that signals more localized guardianship leading to reduced crime compared to regular block party events (such as religious holidays).   In addition, spontaneous block party events may be more inclusive to newer community members which could also increase collective efficacy towards crime prevention.

More generally, the relationships between community vibrancy, collective efficacy and public safety are subtle, nuanced and presumably influenced by many types of neighborhood contexts.  Thus, higher resolution data and measures of community vibrancy and guardianship, such as direct measures of human occupancy and usage of public spaces, are needed for future study.  

\section{Acknowledgements}\label{acknowledgements}

We thank Jon Geeting for providing the block party permit data that was used in our analyses.  

%\newpage

\bibliographystyle{SageH}
\bibliography{references}

\newpage

\setcounter{section}{0}

\begin{center}
{\LARGE {\bf Supplementary Materials for ``Community vibrancy and its relationship with safety in Philadelphia"}}
\end{center}

\bigskip

\section{Spatial Distribution of Community Measures in Philadelphia}

\begin{figure}[!h]
\renewcommand\thefigure{S1}
\centering
\includegraphics[width=0.49\textwidth]{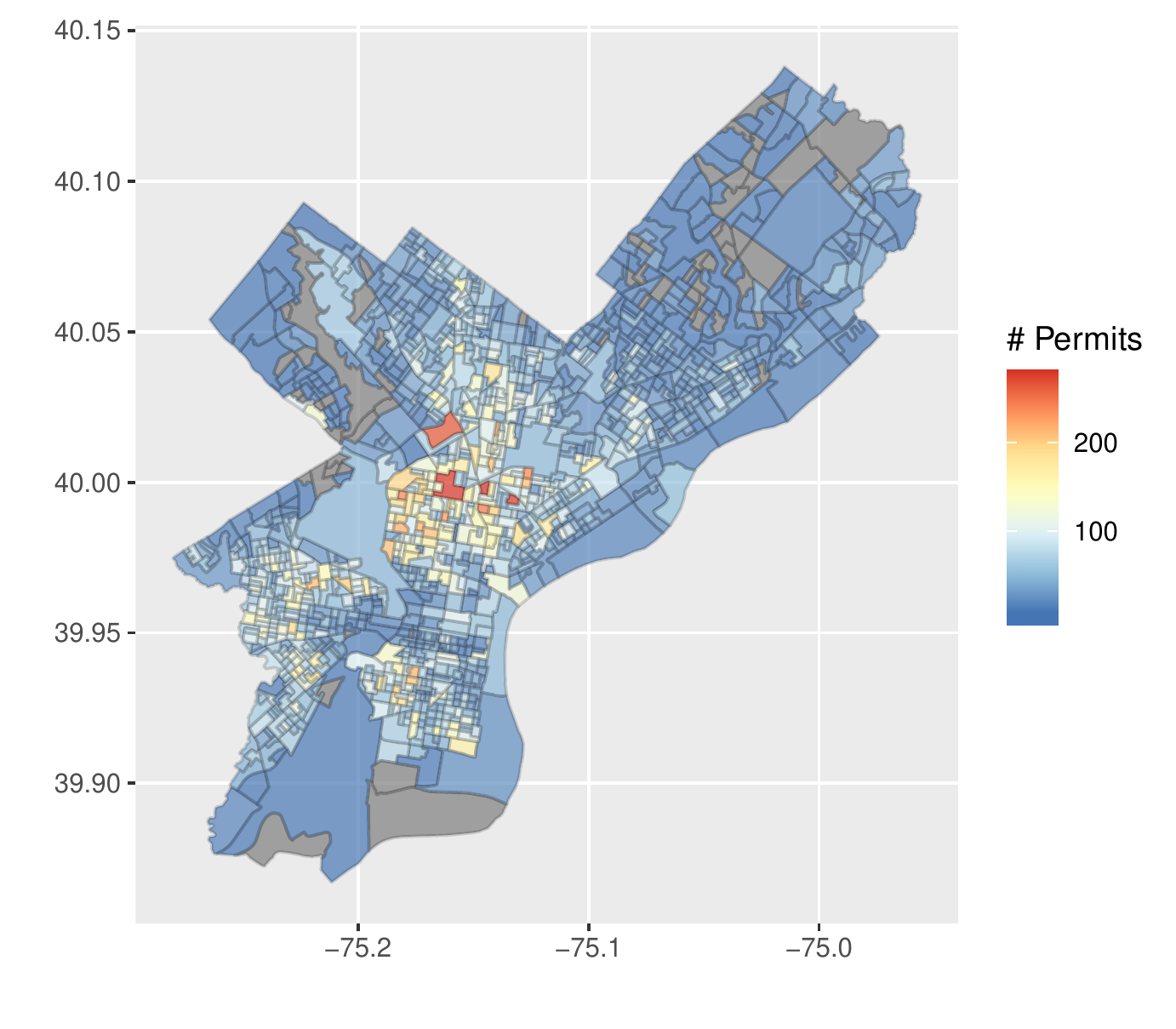} % used to be png
\includegraphics[width=0.49\textwidth]{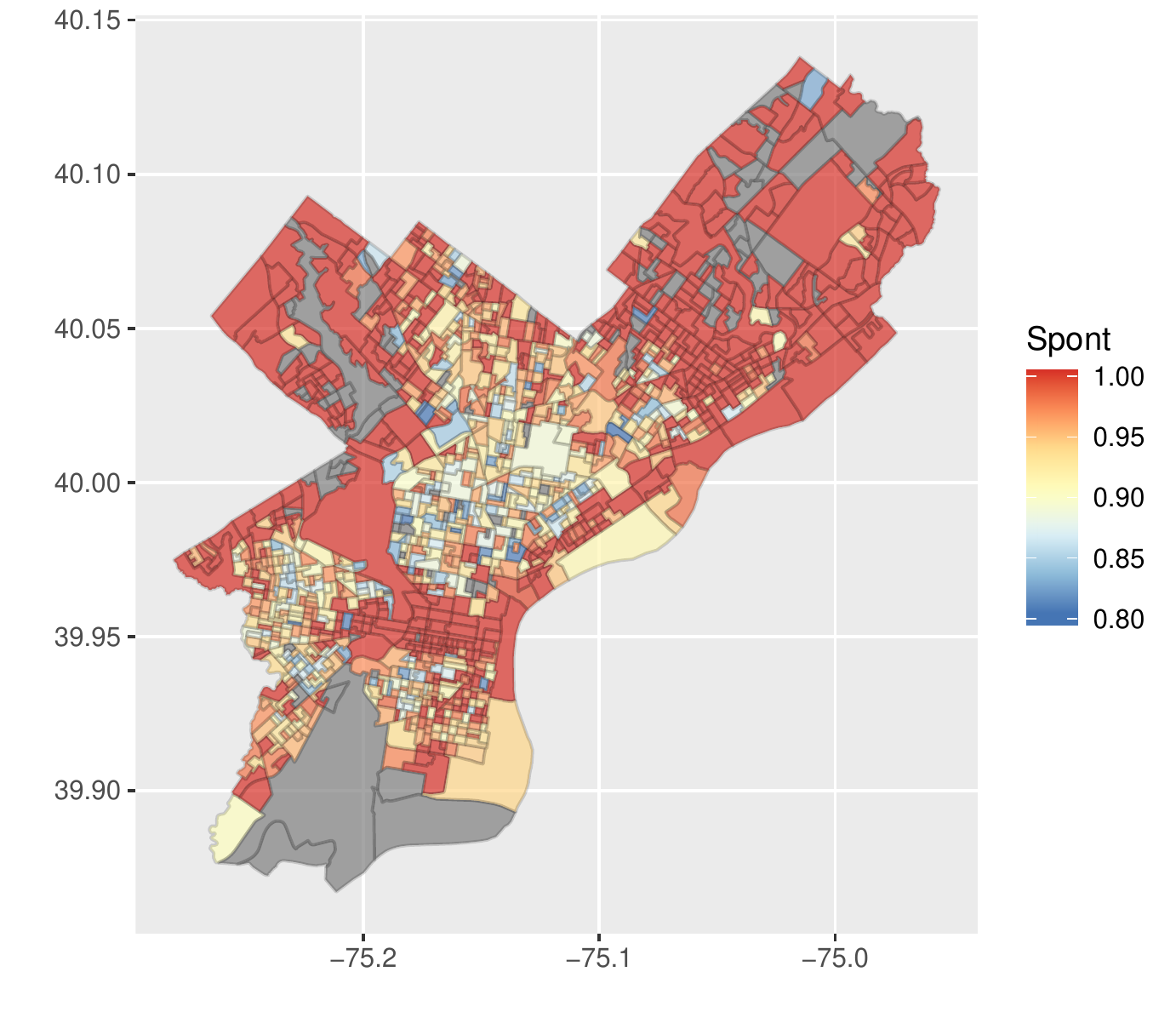} % used to be png
\caption{{\bf Left:} Map of Philadelphia showing the total number of permits per neighborhood.  {\bf Right:} Map of Philadelphia showing the proportion of spontaneous to regular events per neighborhood}
\label{vibrancymeasures}
\end{figure}

In Fig~\ref{vibrancymeasures} (left), we show the total number of block party events within each neighborhood of Philadelphia, aggregated across the entire time span of our data (2006-2016).  We see in Fig~\ref{vibrancymeasures} (left) that neighborhoods that have the largest total number of block party events are in the North Philadelphia area.  West Philadelphia and South Philadelphia also have several neighborhoods with a large total number of block party events, whereas the outlying suburban communities in the Northwest and Northeast parts of the city have relatively few total number of block party events.  

In Fig~\ref{vibrancymeasures} (right), we show the spontaneous within each neighborhood of Philadelphia, based on the total number of spontaneous and regular events across the entire time span of our data (2006-2016). It is interesting to observe that while North Philadelphia contains the neighborhoods with the largest total number of block party events, these North Philadelphia neighborhoods also have a lower spontaneous proportion than the areas of the city that have a smaller number of block party events. Center city and the Northwest and Northeast suburban communities contain the neighborhoods with the highest spontaneous proportions in Philadelphia. 

\section{Monthly Trends in Community Measures in Philadelphia}

\begin{figure}[!h]
\renewcommand\thefigure{S2}
\centering
\includegraphics[width=0.49\textwidth]{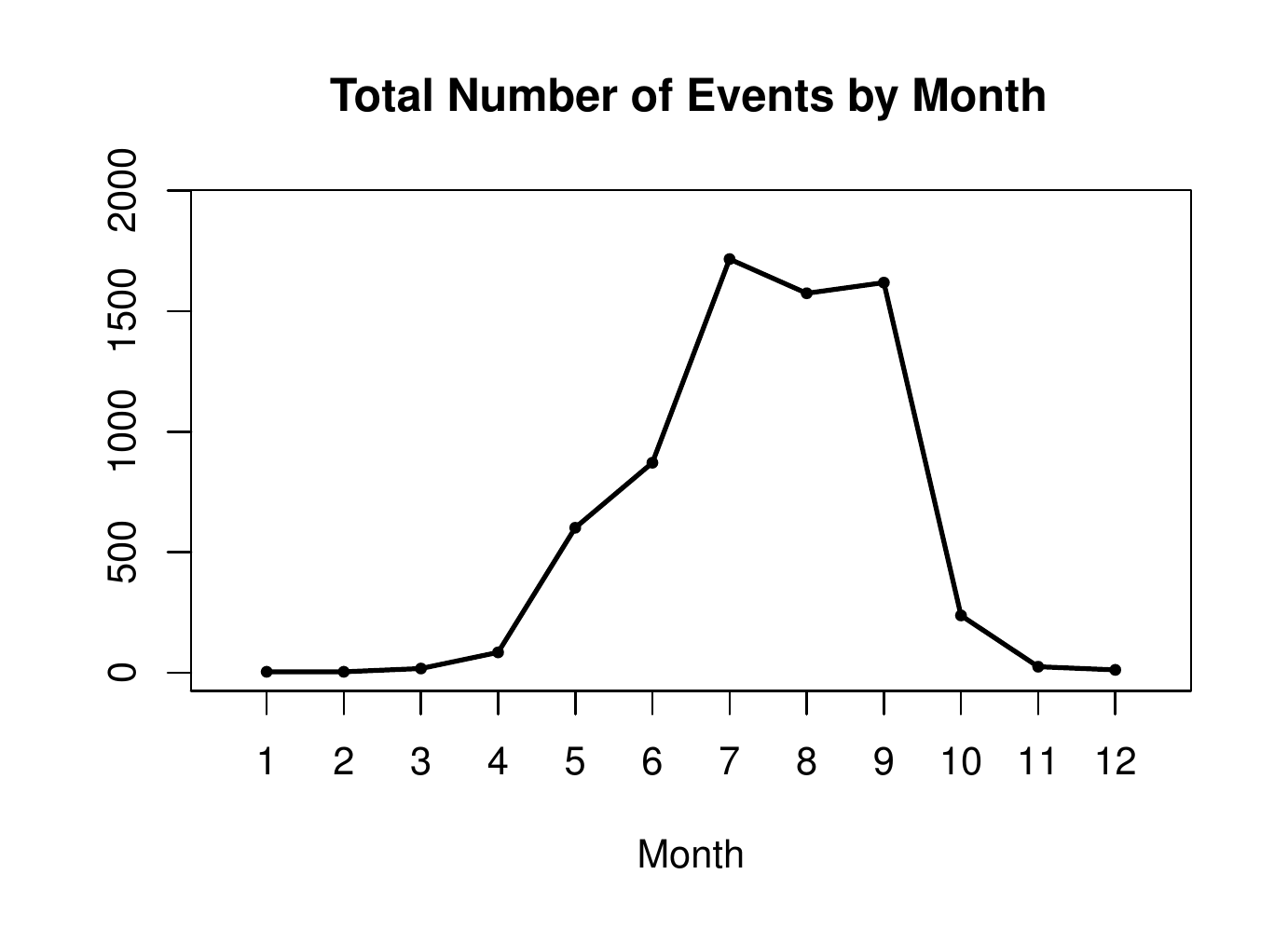}
\includegraphics[width=0.49\textwidth]{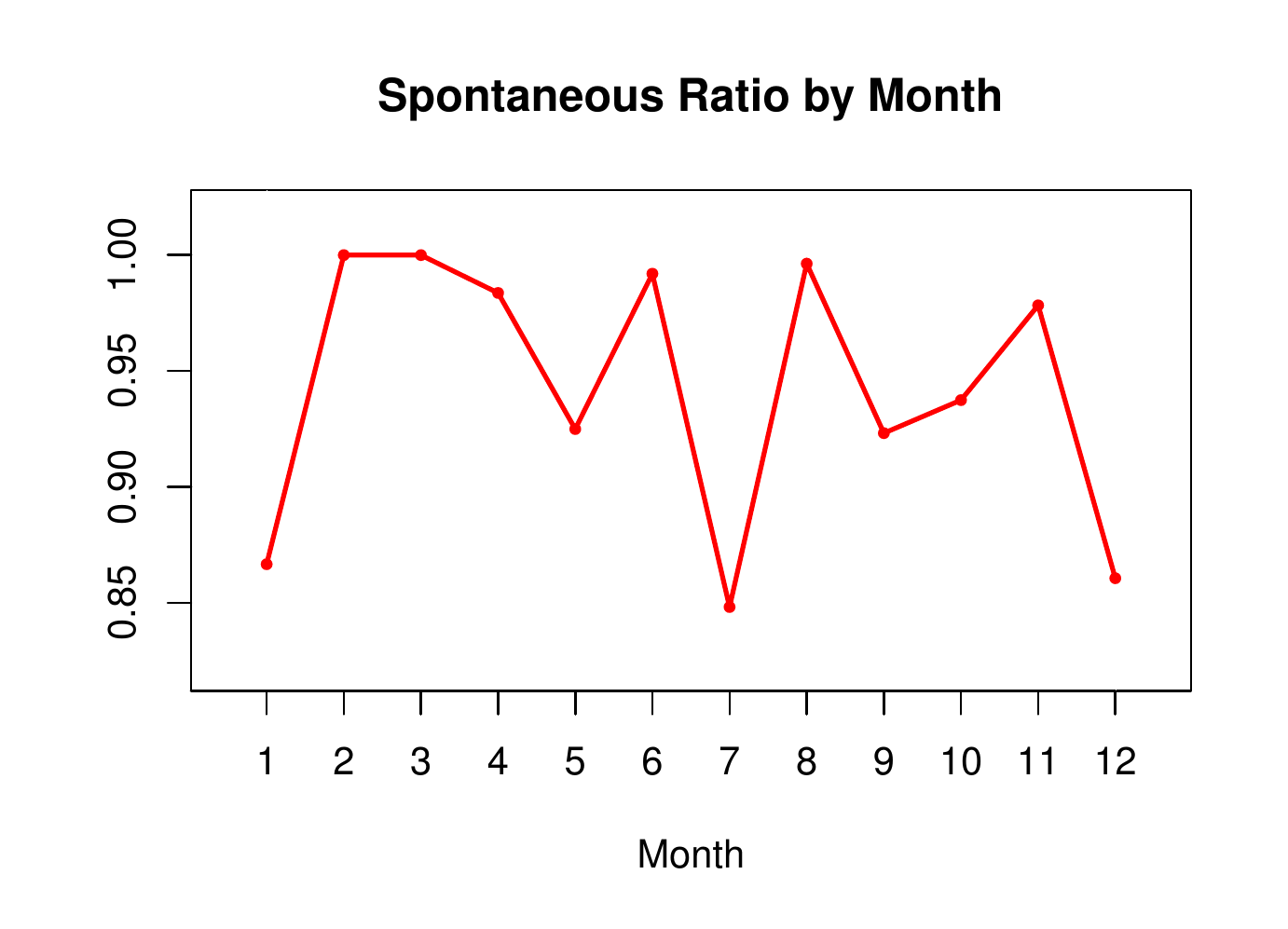}
\caption{Monthly trends in the total number of block party events and the spontaneous proportion of block party events. }
\label{monthlytrends}
\end{figure}

In Fig~\ref{monthlytrends}, we show the variation by month of the total number of block party events and the spontaneous proportion of block party events that we introduced in our main paper.  Note that 2016 is not included in Fig~\ref{monthlytrends} since we only have data for part of that year.

In terms of monthly or seasonal variation, we see a clear trend in the left of Fig~\ref{monthlytrends} for a greater number of block party events during the warmer months from May (5) to September (9).  The spontaneous proportion is lower in May (5), July (7) and September (9) which is due to the prominence of regular holidays (Memorial day, 4th of July, and Labor day) during those months.

\section{Distribution of Total Crimes across Philadelphia Neighborhoods}

\begin{figure}[!h]
\renewcommand\thefigure{S3}
\centering
\includegraphics[width=0.49\textwidth]{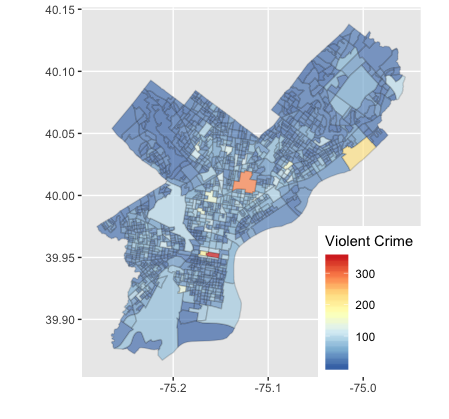}
\includegraphics[width=0.49\textwidth]{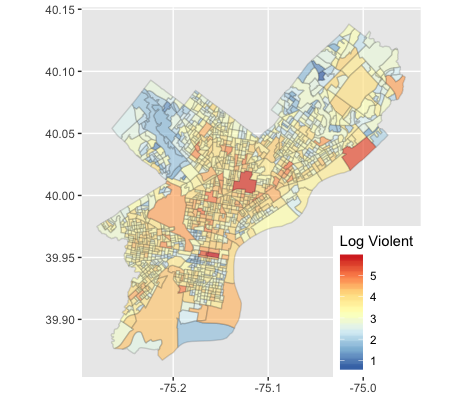}
\caption{Distribution of violent crime over the block groups of Philadelphia. \textbf{Left:} violent crimes per block group, averaged over the years from 2006 to 2015. \textbf{Right:} logarithm of violent crimes per block group, averaged over the years from 2006 to 2015.}
\label{fig:crimemaps}
\end{figure}

Fig~\ref{fig:crimemaps} is a map of the spatial distribution of total crimes per year (averaged over the years from 2006-2015) in Philadelphia, as well as the log transformation of total crimes per year.  

\begin{figure}[!h]
\renewcommand\thefigure{S4}
\centering
\includegraphics[width=0.49\textwidth]{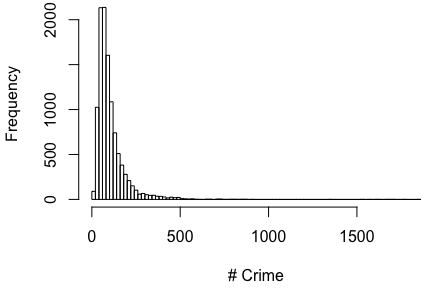}
\includegraphics[width=0.49\textwidth]{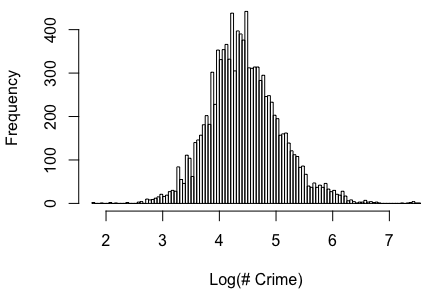}
\caption{{\rm Left}: Distribution of the total number of crimes by US census block group in Philadelphia and {\rm Right}: Distribution of the logarithm of total crimes.}
\label{histcrime}
\end{figure}

Fig~\ref{histcrime} (left) gives the distribution of total crimes over the entire time span across these 1336 neighborhoods.  Since that distribution is highly skewed, we will focus on the log transformation of crime in our analyses which has the more symmetric distribution shown in Fig~\ref{histcrime} (right). 

\section{Correlations between Crime, Community Vibrancy and Other Neighborhood Characteristics}

\begin{figure}[!h]
\renewcommand\thefigure{S5}
\centering
\includegraphics[width=0.7\textwidth]{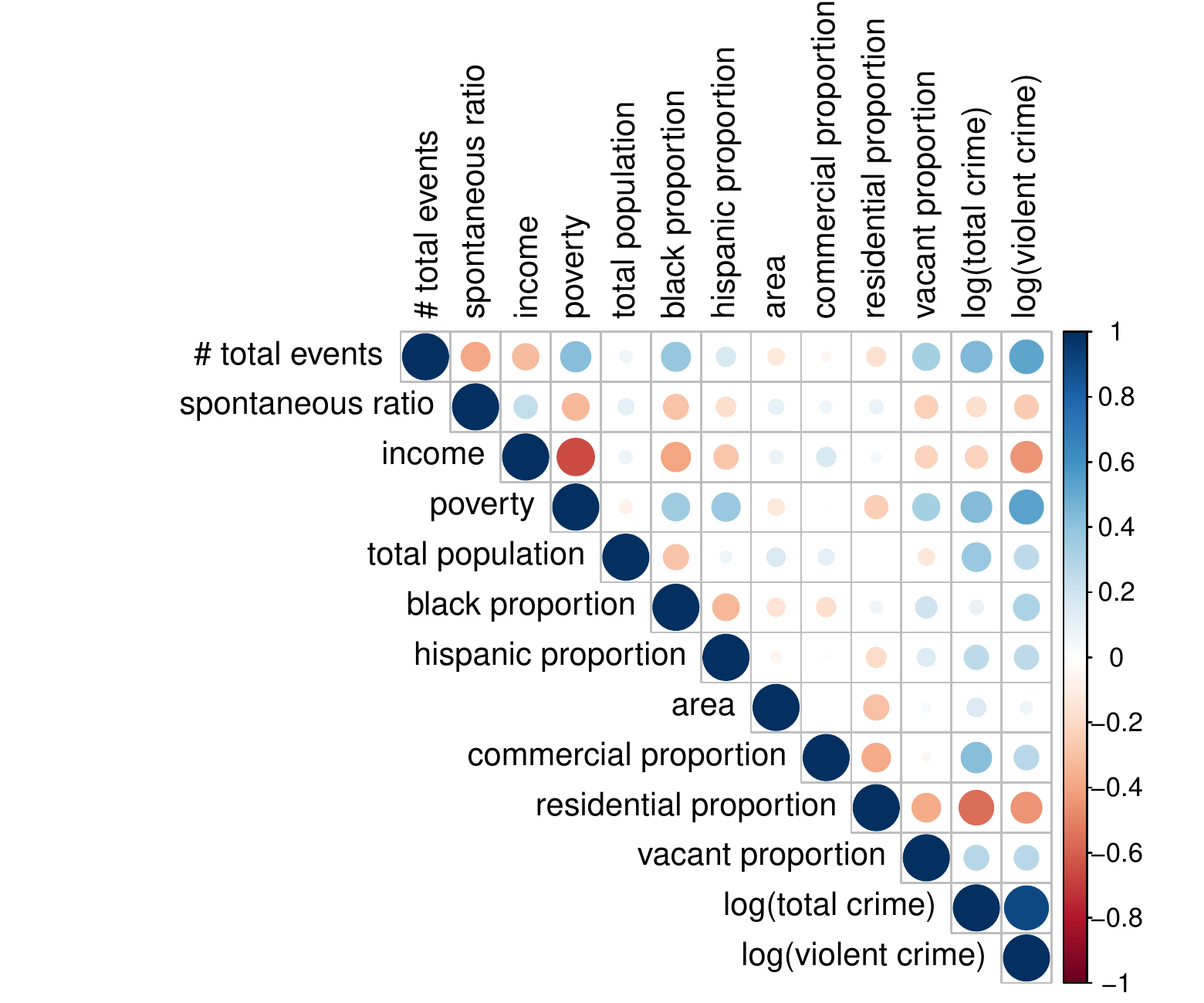}
\caption{Correlations between community vibrancy, demographic, economic, land use and crime measures across all block groups in Philadelphia. Blue indicates a positive correlation, while red reflects a negative correlation. The darker the shade, the larger the magnitude of the correlation.}
\label{fig:corr}
\end{figure}

Fig~\ref{fig:corr} provides correlations between our measures of community vibrancy, several crime measures and the demographic, economic and built environment measures collected for Philadelphia.  

We observe that spontaneous proportion of block parties is not strongly correlated with any of these other neighborhood characteristics.  However, the total number of block party permits is correlated with both economic measures (median income and poverty index) as well as the proportion of black residents in a neighborhood.  We also see that the total number of block party permits is correlated with our measures of crime incidence, and that those crime measures are strongly correlated with several other neighborhood characteristics.

\section{Linear Regression Analysis of Total Crime and Community Vibrancy}

In our main paper, we discuss the results from four different regressions that represent each combination of our two community vibrancy measures and our two regression model specifications, 
\begin{enumerate}
\item Ordinary least squares (OLS) regression of the logarithm of total crime incidence $\log (y_{i})$ on the number of total events $C_i$ and other neighborhood characteristics $\X_i$
\item Ordinary least squares (OLS) regression of the logarithm of total crime incidence $\log (y_{i})$ on the spontaneous proportion $C_i$ and other neighborhood characteristics $\X_i$
\item Negative binomial regression of total crime incidence $\log (y_{i})$ on the number of total events $C_i$ and other neighborhood characteristics $\X_i$
\item Negative binomial regression of total crime incidence $\log (y_{i})$ on the spontaneous proportion $C_i$ and other neighborhood characteristics $\X_i$
\end{enumerate}

Table~\ref{reg1} displays the parameter estimates and model fit statistics for the four regression models outlined above.  
\begin{table}[!h] 
\renewcommand\thetable{S1}
\centering 
\footnotesize
  \caption{Regression model summaries for four different models with the number of total crimes as the outcome variable} 
  \label{reg1} 
\begin{tabular}{@{\extracolsep{5pt}}lcccc} 
\\[-1.8ex]\hline 
\hline \\[-1.8ex] 
& \multicolumn{4}{c}{\textit{Dependent variable:}} \\ 
\cline{2-5} 
\\[-1.8ex] & \multicolumn{2}{c}{Log number of total crimes} & \multicolumn{2}{c}{Number of total crimes} \\ 
\\[-1.8ex] & \multicolumn{2}{c}{\textit{OLS}} & \multicolumn{2}{c}{\textit{negative}} \\ 
& \multicolumn{2}{c}{\textit{}} & \multicolumn{2}{c}{\textit{binomial}} \\ 
\\[-1.8ex] & (1) & (2) & (3) & (4)\\ 
\hline \\[-1.8ex] 
\# Permits & 0.002$^{***}$ (0.0003) & & 0.002$^{***}$ (0.0003) & \\ 
Spontaneity ratio & & $-$0.281$^{+}$ (0.158) & & $-$0.229 (0.156) \\ 
Log income & 0.014 (0.030) & 0.016 (0.031) & 0.026 (0.030) & 0.028 (0.031) \\ 
Poverty & 0.681$^{***}$ (0.090) & 0.751$^{***}$ (0.093) & 0.733$^{***}$ (0.089) & 0.796$^{***}$ (0.092) \\ 
Log population & 0.633$^{***}$ (0.027) & 0.713$^{***}$ (0.027) & 0.629$^{***}$ (0.027) & 0.708$^{***}$ (0.026) \\ 
Black & 0.310$^{***}$ (0.037) & 0.430$^{***}$ (0.036) & 0.296$^{***}$ (0.037) & 0.411$^{***}$ (0.036) \\ 
Hispanic & 0.399$^{***}$ (0.070) & 0.498$^{***}$ (0.072) & 0.407$^{***}$ (0.069) & 0.493$^{***}$ (0.071) \\ 
Area ($10^6$) & 0.177$^{***}$ (0.023) & 0.162$^{***}$ (0.024) & 0.203$^{***}$ (0.023) & 0.193$^{***}$ (0.024) \\ 
Commercial & 2.700$^{***}$ (0.172) & 2.532$^{***}$ (0.177) & 2.739$^{***}$ (0.171) & 2.553$^{***}$ (0.175) \\ 
Residential & $-$1.236$^{***}$ (0.103) & $-$1.437$^{***}$ (0.105) & $-$1.357$^{***}$ (0.102) & $-$1.587$^{***}$ (0.103) \\ 
Vacant & 0.282 (0.208) & 0.479$^{*}$ (0.214) & 0.194 (0.206) & 0.375$^{+}$ (0.212) \\ 
Transportation & $-$0.024$^{**}$ (0.009) & $-$0.030$^{**}$ (0.010) & $-$0.023$^{*}$ (0.009) & $-$0.029$^{**}$ (0.009) \\ 
Industrial & $-$0.017 (0.160) & $-$0.224 (0.164) & $-$0.072 (0.159) & $-$0.322$^{*}$ (0.162) \\ 
Park & $-$0.819$^{***}$ (0.165) & $-$0.995$^{***}$ (0.169) & $-$0.853$^{***}$ (0.164) & $-$1.065$^{***}$ (0.167) \\ 
Civic & 0.293$^{*}$ (0.137) & 0.107 (0.141) & 0.292$^{*}$ (0.136) & 0.097 (0.139) \\ 
Constant & 2.714$^{***}$ (0.384) & 2.561$^{***}$ (0.427) & 2.725$^{***}$ (0.382) & 2.537$^{***}$ (0.422) \\ 
\hline \\[-1.8ex] 
Observations & 1,265 & 1,265 & 1,265 & 1,265 \\ 
R$^{2}$ & 0.718 & 0.697 & & \\ 
Adjusted R$^{2}$ & 0.715 & 0.694 & & \\ 
%Log Likelihood & & & $-$9,828.486 & $-$9,867.931 \\ 
%$\theta$ & & & 9.849$^{***}$ (0.387) & 9.270$^{***}$ (0.364) \\ 
Akaike Inf. Crit. & & & 19,686.970 & 19,765.860 \\ 
%Residual Std. Error (df = 1250) & 0.322 & 0.333 & & \\ 
%F Statistic (df = 14; 1250) & 226.995$^{***}$ & 205.469$^{***}$ & & \\ 
RMSE & 0.3198 & 0.3313 & 0.3403 & 0.3488 \\ 
\hline 
\hline \\[-1.8ex] 
\textit{Note:} & \multicolumn{4}{r}{+p$<$0.01; $^{*}$ p $<$0.05; $^{**}$p$<$0.01; $^{***}$p$<$0.001} \\ 
\end{tabular} 
\end{table}

We see in Table~\ref{reg1} that OLS regression models are a better fit to the data than the negative binomial regression models in terms of root mean square error (RMSE).  We also see that the number of total permits is significantly positively associated with total crimes (models 1 and 3), whereas the spontaneous proportion is non-significantly negatively associated with total crimes (models 2 and 4).  

We also see that most neighborhood characteristics have significant partial effects, which suggests that each of these economic, demographic and land use characteristics have an association with crime, even after accounting for the other characteristics included in the model.  Higher levels of poverty and larger commercial proportions are associated with higher levels of total crime in each of the four models, whereas higher proportions of park space and residential land use are associated with lower levels of total crime.    

\subsection{Regression Analysis of Trends over Time}

In our main paper, we used regression models to explore the neighborhoods factors that are predictive of whether or not a neighborhood has a significant trend over time in either crime or our measures of community vibrancy.   Specifically, we fit the four different logistic regression models enumerated below: 
\begin{enumerate}
\item Logistic regression with significantly increasing trend in community (or not) as the outcome and neighborhood characteristics $\X_i$ (including indicators of trends in crime) as the predictors 
\item Logistic regression with significantly decreasing trend in community (or not) as the outcome and neighborhood characteristics $\X_i$ (including indicators of trends in crime) as the predictors
\item Logistic regression with significantly increasing trend in crime (or not) as the outcome and neighborhood characteristics $\X_i$ (including indicators of trends in community) as the predictors
\item Logistic regression with significantly decreasing trend in crime (or not) as the outcome and neighborhood characteristics $\X_i$ (including indicators of trends in community) as the predictors
\end{enumerate}

Table \ref{trendtab} displays the parameter estimates and model fit statistics for the four logistic regression models listed above, where we use the number of block party permits as our measure of community.   We see that log income is a strong predictor of significantly increasing trends in block party permits (model 1) and that vacant proportion is a strong predictor of significantly decreasing trends in block party permits (model 2).  We also see that industrial land use is a strong predictor of a significantly increasing trend in crime (model 3) and that the Hispanic proportion is a strong predictor of a significantly decreasing trend in crime (model 4).  It is worth noting that for predicting significantly increasing trends in either block party permits or crimes, there are so few cases of either of those two outcomes ($n=94$ in model 1 and $n=38$ in model 3) which which gives us limited power to detect strong associations.   

\begin{table}[!h]  
\renewcommand\thetable{S2}
\centering
\small
  \caption{Logistic regression model results for predicting neighborhoods with different types of significant trends over time} 
  \label{trendtab} 
\begin{tabular}{@{\extracolsep{5pt}}lcccc} 
\\[-1.8ex]\hline 
\hline \\[-1.8ex] 
& \multicolumn{4}{c}{\textit{Dependent variable:}} \\ 
\cline{2-5} 
\\[-1.8ex] & Permits $+$ & Permits $-$ & Crimes $+$ & Crimes $-$ \\ 
\\[-1.8ex] & (1) & (2) & (3) & (4)\\ 
\hline \\[-1.8ex] 
Log income & 0.087$^{***}$ (0.024) & 0.001 (0.032) & $-$0.010 (0.016) & 0.110$^{*}$ (0.046) \\ 
Poverty & 0.002 (0.071) & 0.148 (0.097) & 0.034 (0.047) & 0.034 (0.137) \\ 
Log population & 0.015 (0.021) & 0.033 (0.028) & 0.002 (0.014) & 0.063 (0.040) \\ 
Black & $-$0.024 (0.027) & 0.018 (0.037) & 0.036$^{*}$ (0.018) & $-$0.058 (0.052) \\ 
Hispanic & 0.019 (0.055) & 0.100 (0.074) & 0.001 (0.036) & 0.328$^{**}$ (0.105) \\ 
Area ($10^6$) & $-$0.013 (0.018) & $-$0.010 (0.025) & 0.006 (0.012) & $-$0.040 (0.035) \\ 
Commercial & $-$0.013 (0.136) & $-$0.215 (0.186) & 0.222$^{*}$ (0.089) & $-$0.224 (0.261) \\ 
Residential & $-$0.110 (0.080) & 0.111 (0.109) & 0.022 (0.053) & $-$0.271$^{+}$ (0.154) \\ 
Vacant & $-$0.242 (0.164) & 0.723$^{**}$ (0.223) & 0.137 (0.108) & 0.521$^{+}$ (0.316) \\ 
Transportation & $-$0.003 (0.007) & $-$0.006 (0.010) & 0.010$^{*}$ (0.005) & $-$0.009 (0.014) \\ 
Industrial & $-$0.165 (0.126) & $-$0.067 (0.172) & 0.256$^{**}$ (0.083) & $-$0.109 (0.242) \\ 
Park & 0.024 (0.130) & $-$0.030 (0.177) & $-$0.029 (0.085) & $-$0.072 (0.250) \\ 
Civic & 0.179$^{+}$ (0.108) & 0.109 (0.147) & $-$0.038 (0.071) & 0.127 (0.208) \\ 
Crimes $+$ & 0.024 (0.044) & $-$0.053 (0.059) & & \\ 
Crimes $-$ & 0.003 (0.015) & 0.012 (0.020) & & \\ 
Permits $+$ & & & 0.008 (0.019) & 0.012 (0.055) \\ 
Permits $-$ & & & $-$0.013 (0.014) & 0.031 (0.040) \\ 
Constant & $-$0.818$^{**}$ (0.304) & $-$0.230 (0.413) & 0.048 (0.200) & $-$0.970$^{+}$ (0.584) \\ 
\hline \\[-1.8ex] 
Outcome = 1 & 94 & 184 & 38 & 589 \\
Observations & 1,265 & 1,265 & 1,265 & 1,265 \\ 
Log Likelihood & $-$62.720 & $-$452.367 & 468.369 & $-$888.767 \\ 
Akaike Inf. Crit. & 157.439 & 936.734 & $-$904.738 & 1,809.533 \\ 
\hline 
\hline \\[-1.8ex] 
\textit{Note:} & \multicolumn{4}{r}{+p$<$0.01; $^{*}$ p $<$0.05; $^{**}$p$<$0.01; $^{***}$p$<$0.001} \\ 
\end{tabular} 
\end{table}

Table \ref{trendtab2} displays the parameter estimates and model fit statistics for another four logistic regression models in which we use the spontaneity proportion as our measure of community.  We find that several factors are significantly associated with a positive trend over time in spontaneity proportion, e.g. poverty, population size, proportions of Black and Hispanic residents, as well as residential and vacant land use (column 1). On the other hand, the only significant predictors of the much smaller set of neighborhoods with a negative trend in spontaneous proportion are the proportion of Black residents and residential and civic land uses (column 2).    There is no significant association between trends over time in crime and trends over time in the spontaneous proportion

\begin{table}[!h] \centering 
\renewcommand\thetable{S3}
\small
  \caption{Logistic regression model results for predicting neighborhoods with different types of significant trends over time} 
  \label{trendtab2} 
\begin{tabular}{@{\extracolsep{5pt}}lcccc} 
\\[-1.8ex]\hline 
\hline \\[-1.8ex] 
& \multicolumn{4}{c}{\textit{Dependent variable:}} \\ 
\cline{2-5} 
\\[-1.8ex] & Spont $+$ & Spont $-$ & Crimes $+$ & Crimes $-$ \\ 
\\[-1.8ex] & (1) & (2) & (3) & (4)\\ 
\hline \\[-1.8ex] 
Log income & 0.041 (0.037) & 0.016 (0.013) & $-$0.009 (0.016) & 0.109$^{*}$ (0.046) \\ 
Poverty & 0.312$^{**}$ (0.112) & 0.001 (0.040) & 0.033 (0.047) & 0.024 (0.137) \\ 
Log population & 0.085$^{**}$ (0.032) & 0.009 (0.011) & 0.002 (0.014) & 0.060 (0.040) \\ 
Black & 0.289$^{***}$ (0.043) & $-$0.041$^{**}$ (0.015) & 0.036$^{*}$ (0.018) & $-$0.069 (0.053) \\ 
Hispanic & 0.347$^{***}$ (0.086) & $-$0.026 (0.030) & 0.001 (0.036) & 0.317$^{**}$ (0.105) \\ 
Area ($10^6$) & $-$0.055$^{+}$ (0.029) & 0.010 (0.010) & 0.006 (0.012) & $-$0.038 (0.035) \\ 
Commercial & $-$0.448$^{*}$ (0.214) & 0.023 (0.076) & 0.224$^{*}$ (0.090) & $-$0.212 (0.262) \\ 
Residential & $-$0.335$^{**}$ (0.126) & 0.168$^{***}$ (0.045) & 0.020 (0.053) & $-$0.260$^{+}$ (0.155) \\ 
Vacant & 0.687$^{**}$ (0.258) & 0.211$^{*}$ (0.092) & 0.128 (0.108) & 0.501 (0.316) \\ 
Transportation & 0.005 (0.011) & 0.002 (0.004) & 0.010$^{*}$ (0.005) & $-$0.010 (0.014) \\ 
Industrial & $-$0.281 (0.198) & 0.113 (0.071) & 0.255$^{**}$ (0.083) & $-$0.104 (0.242) \\ 
Park & $-$0.052 (0.204) & 0.088 (0.073) & $-$0.029 (0.085) & $-$0.073 (0.250) \\ 
Civic & $-$0.191 (0.169) & 0.211$^{***}$ (0.060) & $-$0.038 (0.071) & 0.133 (0.208) \\ 
Crimes $+$ & 0.001 (0.068) & $-$0.001 (0.024) & & \\ 
Crimes $-$ & 0.030 (0.023) & 0.003 (0.008) & & \\ 
Spont $+$ & & & $-$0.002 (0.012) & 0.045 (0.035) \\ 
Spont $-$ & & & $-$0.003 (0.033) & 0.037 (0.097) \\ 
Constant & $-$0.877$^{+}$ (0.476) & $-$0.279 (0.169) & 0.041 (0.200) & $-$0.936 (0.584) \\ 
\hline \\[-1.8ex] 
Outcome = 1 & 313 & 27 & 38 & 589 \\
Observations & 1,265 & 1,265 & 1,265 & 1,265 \\ 
Log Likelihood & $-$632.752 & 675.492 & 467.787 & $-$888.171 \\ 
Akaike Inf. Crit. & 1,297.503 & $-$1,318.984 & $-$903.574 & 1,808.342 \\ 
\hline 
\hline \\[-1.8ex] 
\textit{Note:} & \multicolumn{4}{r}{+p$<$0.01; $^{*}$ p $<$0.05; $^{**}$p$<$0.01; $^{***}$p$<$0.001} \\ 
\end{tabular} 
\end{table}

\newpage 

\section{Additional Aggregate Regression Models}

In our main paper, we compare the results from four different regressions that represent each combination of our two community vibrancy measures and our two regression model specifications, 
\begin{enumerate}
\item Ordinary least squares (OLS) regression of the logarithm of total crime incidence $\log (y_{i})$ on the number of total events $C_i$ and other neighborhood characteristics $\X_i$
\item Ordinary least squares (OLS) regression of the logarithm of total crime incidence $\log (y_{i})$ on the spontaneous proportion $C_i$ and other neighborhood characteristics $\X_i$
\item Negative binomial regression of total crime incidence $\log (y_{i})$ on the number of total events $C_i$ and other neighborhood characteristics $\X_i$
\item Negative binomial regression of total crime incidence $\log (y_{i})$ on the spontaneous proportion $C_i$ and other neighborhood characteristics $\X_i$
\end{enumerate}

In this section, we provide results from a similar set of regressions but with (a) just violent crimes, (b) just non-violent crimes or (c) just vice crimes as the outcome variable.    Tables \ref{reg2}, \ref{reg3}, and \ref{reg4} displays the parameter estimates and model fit statistics for the four regression models listed above but with violent crime, non-violent crime and vice crime as the outcome variable, respectively.  

We generally observe similar results in Tables~\ref{reg2}-\ref{reg4} to Table~\ref{reg1} in our main paper where total crimes was the outcome variable.   The partial effects for most neighborhood characteristics are significant, suggesting that each of these economic, demographic and land use characteristics are associated with violent, non-violent and vice crime.   Similar to the models for total crime in our main paper, we see that the number of total permits is significantly positively associated with violent, non-violent and vice crimes, whereas the spontaneous proportion is negatively associated with violent, non-violent and vice crimes (though only significant for vice crimes). 

\begin{table}[!h] 
\renewcommand\thetable{S4}
\centering 
\footnotesize
  \caption{Results from linear regression models with the number of violent crimes as the outcome variable} 
  \label{reg2} 
\begin{tabular}{@{\extracolsep{5pt}}lcccc} 
\\[-1.8ex]\hline 
\hline \\[-1.8ex] 
& \multicolumn{4}{c}{\textit{Dependent variable:}} \\ 
\cline{2-5} 
\\[-1.8ex] & \multicolumn{2}{c}{Log number of violent crimes} & \multicolumn{2}{c}{Number of Violent Crimes} \\ 
\\[-1.8ex] & \multicolumn{2}{c}{\textit{OLS}} & \multicolumn{2}{c}{\textit{negative}} \\ 
& \multicolumn{2}{c}{\textit{}} & \multicolumn{2}{c}{\textit{binomial}} \\ 
\\[-1.8ex] & (1) & (2) & (3) & (4)\\ 
\hline \\[-1.8ex] 
\# Permits & 0.003$^{***}$ (0.0003) & & 0.002$^{***}$ (0.0003) & \\ 
Spontaneous ratio & & $-$0.364$^{*}$ (0.180) & & $-$0.267 (0.176) \\ 
Log income & $-$0.190$^{***}$ (0.034) & $-$0.188$^{***}$ (0.035) & $-$0.189$^{***}$ (0.033) & $-$0.191$^{***}$ (0.034) \\ 
Poverty & 0.431$^{***}$ (0.103) & 0.504$^{***}$ (0.106) & 0.525$^{***}$ (0.101) & 0.585$^{***}$ (0.103) \\ 
Log population & 0.626$^{***}$ (0.031) & 0.711$^{***}$ (0.030) & 0.615$^{***}$ (0.030) & 0.697$^{***}$ (0.030) \\ 
Black & 0.683$^{***}$ (0.042) & 0.809$^{***}$ (0.041) & 0.626$^{***}$ (0.041) & 0.741$^{***}$ (0.040) \\ 
Hispanic & 0.625$^{***}$ (0.079) & 0.727$^{***}$ (0.082) & 0.559$^{***}$ (0.078) & 0.638$^{***}$ (0.080) \\ 
Area ($10^6$) & 0.142$^{***}$ (0.026) & 0.126$^{***}$ (0.027) & 0.195$^{***}$ (0.026) & 0.180$^{***}$ (0.026) \\ 
Commercial & 2.262$^{***}$ (0.197) & 2.083$^{***}$ (0.202) & 2.233$^{***}$ (0.192) & 2.028$^{***}$ (0.196) \\ 
Residential & $-$1.145$^{***}$ (0.118) & $-$1.360$^{***}$ (0.119) & $-$1.198$^{***}$ (0.116) & $-$1.451$^{***}$ (0.116) \\ 
Vacant & $-$0.270 (0.237) & $-$0.065 (0.244) & $-$0.435$^{+}$ (0.232) & $-$0.282 (0.238) \\ 
Transportation  & $-$0.062$^{***}$ (0.011) & $-$0.068$^{***}$ (0.011) & $-$0.058$^{***}$ (0.010) & $-$0.064$^{***}$ (0.011) \\ 
Industrial & $-$0.140 (0.183) & $-$0.361$^{+}$ (0.187) & $-$0.213 (0.179) & $-$0.480$^{**}$ (0.182) \\ 
Park  & $-$0.838$^{***}$ (0.188) & $-$1.026$^{***}$ (0.193) & $-$0.841$^{***}$ (0.185) & $-$1.064$^{***}$ (0.189) \\ 
Civic & 0.491$^{**}$ (0.157) & 0.291$^{+}$ (0.160) & 0.698$^{***}$ (0.153) & 0.481$^{**}$ (0.156) \\ 
Constant & 2.811$^{***}$ (0.439) & 2.709$^{***}$ (0.485) & 2.967$^{***}$ (0.431) & 2.865$^{***}$ (0.474) \\ 
\hline \\[-1.8ex] 
Observations & 1,265 & 1,265 & 1,265 & 1,265 \\ 
R$^{2}$ & 0.699 & 0.680 & & \\ 
Adjusted R$^{2}$ & 0.696 & 0.677 & & \\ 
%Log Likelihood & & & $-$7,640.842 & $-$7,674.265 \\ 
%$\theta$ & & & 7.911$^{***}$ (0.318) & 7.504$^{***}$ (0.301) \\ 
Akaike Inf. Crit. & & & 15,311.680 & 15,378.530 \\ 
%Residual Std. Error (df = 1250) & 0.368 & 0.379 & & \\ 
%F Statistic (df = 14; 1250) & 207.836$^{***}$ & 190.101$^{***}$ & & \\ 
RMSE & 0.3653 & 0.3767 & 0.3863 & 0.3937 \\
\hline 
\hline \\[-1.8ex] 
\textit{Note:} & \multicolumn{4}{r}{+p$<$0.01; $^{*}$ p $<$0.05; $^{**}$p$<$0.01; $^{***}$p$<$0.001} \\ 
\end{tabular} 
\end{table}

\begin{table}[!h] 
\renewcommand\thetable{S5}
\centering 
\footnotesize
  \caption{Results from linear regression models with the number of non-violent crimes as the outcome variable} 
  \label{reg3} 
\begin{tabular}{@{\extracolsep{5pt}}lcccc} 
\\[-1.8ex]\hline 
\hline \\[-1.8ex] 
& \multicolumn{4}{c}{\textit{Dependent variable:}} \\ 
\cline{2-5} 
\\[-1.8ex] & \multicolumn{2}{c}{Log number of non-violent crimes} & \multicolumn{2}{c}{Number of non-violent crimes} \\ 
\\[-1.8ex] & \multicolumn{2}{c}{\textit{OLS}} & \multicolumn{2}{c}{\textit{negative}} \\ 
& \multicolumn{2}{c}{\textit{}} & \multicolumn{2}{c}{\textit{binomial}} \\ 
\\[-1.8ex] & (1) & (2) & (3) & (4)\\ 
\hline \\[-1.8ex] 
\# Permits & 0.002$^{***}$ (0.0002) & & 0.001$^{***}$ (0.0002) & \\ 
Spontaneous ratio & & $-$0.187 (0.151) & & $-$0.127 (0.151) \\ 
Log income & 0.085$^{**}$ (0.029) & 0.086$^{**}$ (0.030) & 0.097$^{***}$ (0.029) & 0.099$^{***}$ (0.029) \\ 
Poverty & 0.579$^{***}$ (0.088) & 0.622$^{***}$ (0.089) & 0.609$^{***}$ (0.088) & 0.649$^{***}$ (0.089) \\ 
Log population & 0.695$^{***}$ (0.027) & 0.744$^{***}$ (0.026) & 0.691$^{***}$ (0.026) & 0.738$^{***}$ (0.026) \\ 
Black & 0.044 (0.036) & 0.118$^{***}$ (0.035) & 0.023 (0.036) & 0.092$^{**}$ (0.035) \\ 
Hispanic & 0.054 (0.068) & 0.115$^{+}$ (0.069) & 0.020 (0.068) & 0.072 (0.068) \\ 
Area ($10^6$) & 0.166$^{***}$ (0.023) & 0.157$^{***}$ (0.023) & 0.193$^{***}$ (0.022) & 0.188$^{***}$ (0.023) \\ 
Commercial  & 2.839$^{***}$ (0.168) & 2.735$^{***}$ (0.170) & 2.913$^{***}$ (0.167) & 2.803$^{***}$ (0.168) \\ 
Residential & $-$1.282$^{***}$ (0.101) & $-$1.406$^{***}$ (0.100) & $-$1.415$^{***}$ (0.100) & $-$1.550$^{***}$ (0.099) \\ 
Vacant & 0.238 (0.203) & 0.359$^{+}$ (0.205) & 0.197 (0.202) & 0.308 (0.204) \\ 
Transportation  & 0.001 (0.009) & $-$0.003 (0.009) & 0.002 (0.009) & $-$0.001 (0.009) \\ 
Industrial & 0.258$^{+}$ (0.156) & 0.130 (0.157) & 0.227 (0.156) & 0.080 (0.156) \\ 
Park  & $-$0.732$^{***}$ (0.161) & $-$0.841$^{***}$ (0.162) & $-$0.776$^{***}$ (0.161) & $-$0.906$^{***}$ (0.161) \\ 
Civic & 0.148 (0.134) & 0.032 (0.135) & 0.162 (0.133) & 0.050 (0.134) \\ 
Constant & 0.606 (0.376) & 0.523 (0.408) & 0.620$^{+}$ (0.375) & 0.487 (0.406) \\ 
\hline \\[-1.8ex] 
Observations & 1,265 & 1,265 & 1,265 & 1,265 \\ 
R$^{2}$ & 0.705 & 0.696 & & \\ 
Adjusted R$^{2}$ & 0.701 & 0.693 & & \\ 
%Log Likelihood & & & $-$8,239.780 & $-$8,255.090 \\ 
%$\theta$ & & & 10.371$^{***}$ (0.414) & 10.125$^{***}$ (0.403) \\ 
Akaike Inf. Crit. & & & 16,509.560 & 16,540.180 \\ 
%Residual Std. Error (df = 1250) & 0.315 & 0.319 & & \\ 
%F Statistic (df = 14; 1250) & 213.060$^{***}$ & 204.473$^{***}$ & & \\ 
RMSE & 0.3127 & 0.3172 & 0.3379 & 0.3413 \\
\hline 
\hline \\[-1.8ex] 
\textit{Note:} & \multicolumn{4}{r}{+p$<$0.01; $^{*}$ p $<$0.05; $^{**}$p$<$0.01; $^{***}$p$<$0.001} \\ 
\end{tabular} 
\end{table}

\begin{table}[!h] 
\renewcommand\thetable{S6}
\centering 
\footnotesize
  \caption{Results from linear regression models with the number of vice crimes as the outcome variable} 
  \label{reg4} 
\begin{tabular}{@{\extracolsep{5pt}}lcccc} 
\\[-1.8ex]\hline 
\hline \\[-1.8ex] 
& \multicolumn{4}{c}{\textit{Dependent variable:}} \\ 
\cline{2-5} 
\\[-1.8ex] & \multicolumn{2}{c}{Log number of vice crimes} & \multicolumn{2}{c}{Number of vice crimes} \\ 
\\[-1.8ex] & \multicolumn{2}{c}{\textit{OLS}} & \multicolumn{2}{c}{\textit{negative}} \\ 
& \multicolumn{2}{c}{\textit{}} & \multicolumn{2}{c}{\textit{binomial}} \\ 
\\[-1.8ex] & (1) & (2) & (3) & (4)\\ 
\hline \\[-1.8ex] 
\# Permits & 0.008$^{***}$ (0.001) & & 0.007$^{***}$ (0.001) & \\ 
Spontaneous ratio & & $-$1.461$^{**}$ (0.538) & & $-$0.991$^{*}$ (0.387) \\ 
Log income & $-$0.343$^{***}$ (0.102) & $-$0.339$^{**}$ (0.105) & $-$0.164$^{*}$ (0.073) & $-$0.144$^{+}$ (0.076) \\ 
Poverty & 1.023$^{***}$ (0.308) & 1.226$^{***}$ (0.316) & 1.510$^{***}$ (0.220) & 1.638$^{***}$ (0.228) \\ 
Log population & 0.403$^{***}$ (0.093) & 0.659$^{***}$ (0.091) & 0.332$^{***}$ (0.066) & 0.599$^{***}$ (0.066) \\ 
Black & 1.716$^{***}$ (0.126) & 2.073$^{***}$ (0.123) & 1.391$^{***}$ (0.091) & 1.721$^{***}$ (0.089) \\ 
Hispanic & 1.695$^{***}$ (0.238) & 1.969$^{***}$ (0.244) & 1.924$^{***}$ (0.169) & 2.157$^{***}$ (0.175) \\ 
Area ($10^6$) & 0.031 (0.079) & $-$0.013 (0.081) & 0.140$^{*}$ (0.056) & 0.065 (0.058) \\ 
Commercial & 2.866$^{***}$ (0.588) & 2.344$^{***}$ (0.603) & 2.352$^{***}$ (0.418) & 1.749$^{***}$ (0.432) \\ 
Residential & $-$2.804$^{***}$ (0.352) & $-$3.439$^{***}$ (0.355) & $-$2.268$^{***}$ (0.253) & $-$3.037$^{***}$ (0.257) \\ 
Vacant & $-$0.439 (0.709) & 0.125 (0.729) & $-$0.074 (0.504) & 0.365 (0.521) \\ 
Transportation & $-$0.074$^{*}$ (0.031) & $-$0.095$^{**}$ (0.032) & $-$0.095$^{***}$ (0.022) & $-$0.116$^{***}$ (0.023) \\ 
Industrial & $-$1.721$^{**}$ (0.547) & $-$2.376$^{***}$ (0.558) & $-$1.395$^{***}$ (0.389) & $-$2.274$^{***}$ (0.401) \\ 
Park & $-$1.453$^{**}$ (0.563) & $-$2.006$^{***}$ (0.576) & $-$1.301$^{**}$ (0.404) & $-$1.835$^{***}$ (0.416) \\ 
Civic & $-$0.197 (0.469) & $-$0.789$^{+}$ (0.478) & $-$0.009 (0.334) & $-$0.643$^{+}$ (0.343) \\ 
Constant & 3.751$^{**}$ (1.313) & 3.826$^{**}$ (1.449) & 2.622$^{**}$ (0.942) & 2.116$^{*}$ (1.047) \\ 
\hline \\[-1.8ex] 
Observations & 1,265 & 1,265 & 1,265 & 1,265 \\ 
R$^{2}$ & 0.554 & 0.528 & & \\ 
Adjusted R$^{2}$ & 0.549 & 0.522 & & \\ 
%Log Likelihood & & & $-$6,468.459 & $-$6,524.143 \\ 
%$\theta$ & & & 1.688$^{***}$ (0.066) & 1.552$^{***}$ (0.060) \\ 
Akaike Inf. Crit. & & & 12,966.920 & 13,078.290 \\ 
%Residual Std. Error (df = 1250) & 1.100 & 1.132 & & \\ 
%F Statistic (df = 14; 1250) & 111.124$^{***}$ & 99.722$^{***}$ & & \\ 
RMSE & 1.0932 & 1.1257 & 0.9551 & 0.9951 \\
\hline 
\hline \\[-1.8ex] 
\textit{Note:} & \multicolumn{4}{r}{+p$<$0.01; $^{*}$ p $<$0.05; $^{**}$p$<$0.01; $^{***}$p$<$0.001} \\ 
\end{tabular} 
\end{table}

%\Floatbarrier

\end{document}